\documentclass[referee]{jfm}

\usepackage{amsmath,amssymb}
\usepackage{graphicx,psfrag}
\usepackage{color}

\newcommand{\eql}{\begin{equation}\label}
\newcommand{\eqn}[1]{(\ref{#1})}
\newcommand{\Dx}{\Delta x}
\newcommand{\Dt}{\Delta t}
\newcommand{\CFL}{{\text\bf CFL}}
\newcommand{\xmin}{x_{\text{min}}}
\newcommand{\Nout}{N_{\text{out}}}
\newcommand{\new}[1]{#1}

\begin{document}


\title[Universality in the run-up of shock waves]%
{Universality in the run-up of shock waves to the surface of a star}

\author[C. Gundlach and R. J. LeVeque]{C.\ns G\ls U\ls N\ls D\ls L\ls A\ls C\ls H$^1$ 
\and R.\ns J.\ns L\ls E\ls V\ls E\ls Q\ls U\ls E$^2$ }

\affiliation{$^1$School of Mathematics, University of Southampton,
  Southampton, SO17 1BJ, United Kingdom\\[\affilskip]
 $^2$Department of Applied Mathematics, University of Washington,
Box 352420, Seattle, WA 98195-2420}
\date{11 August 2010}


\begin{abstract}

We investigate the run-up of a shock wave from inside to the surface
of a perfect fluid star in equilibrium and bounded by vacuum. Near the
surface we approximate the fluid motion as plane-symmetric and the
gravitational field as constant. \new{We consider the
  ``hot'' equation of state $P=(\Gamma-1)\rho e$ and its ``cold''
  (fixed entropy, barotropic) form $P=K_0\rho^\Gamma$ (the latter does
  not allow for shock heating).}  We find numerically that the
evolution of generic initial data approaches universal similarity
solutions sufficiently near the surface, and we construct these
similarity solutions explicitly. The two equations of state show very
different behaviour, because shock heating becomes the dominant effect
\new{when it is allowed. In the barotropic} case, the fluid velocity
behind the shock approaches a constant value, while the density behind
the shock approaches a power law \new{in space, as the shock
  approaches the surface}.  In the \new{hot case with shock heating},
the density jumps by a constant factor through the shock, while the
sound speed and fluid velocity behind the shock diverge in a whiplash
effect. We tabulate the similarity exponents as a function of 
\new{the equation of state parameter} $\Gamma$
and the stratification index $n_*$.

\end{abstract}


\maketitle



\section{Introduction}


In the numerical simulation of high-energy astrophysical events, one
is often faced with three linked unknowns: the mathematical model of
the relevant physics (such as an equation of state), the
qualitative features to expect (such as shocks) and a numerical method
that correctly represents these features. This is the case for the
surfaces of compact binaries just before merger.

The full simulation of compact binary mergers in general relativity
has been achieved by a small number of groups in recent years. Recent
papers describing their methods include \cite{BaiottiBNS2008} and
\cite{ShibataBNS2009} for binary neutron star mergers, and
\cite{EtienneBHNS2009} and \cite{ShibataBHNS2010} for black
hole-neutron star mergers, all using high-resolution shock-capturing
(HRSC) methods (see \cite{Font2008} for a review) for simulating the
matter, and \cite{Rosswog2002} and\cite{Janka2010} for binary neutron
star mergers using smoothed particle hydrodynamics (SPH) methods.

Given the complications involved in general relativity and the large
computational requirements in 3-dimensional simulations involving a
range of length scales, in all these simulations the matter is
modelled as a single perfect fluid. \new{The equations of state used range
  from the two simple families that we also consider here, 
  over composite polytropes and tabulated realistic cold equations of
  state to combinations of a tabulated cold component and a simple hot
  component. See \cite{Duez2009} for a review. If we take current
  single fluid models} seriously, the neutron star or stars in these
models have a surface at finite radius where both the density and
pressure reach zero, with vacuum outside. \new{(By contrast, if a star
  has a solid crust, the density is small but finite at the surface,
  and the results of this paper do not apply.)}

Both HRSC and SPH methods have serious difficulties in correctly
modelling the physics in this region. This is essentially because
quantities such as the density and pressure approach zero as powers of
distance to the surface, and hence are not smooth at the surface
itself. This affects in particular the numerical modelling of the
balance between the gravitational field and pressure gradient in a
star in equilibrium.  HRSC methods, such as those cited above, replace
the vacuum exterior by an unphysical low-density ``numerical
atmosphere'' with a numerically modified time evolution. SPH methods
put few particles there. As a consequence, current binary evolution
codes are incapable of correctly representing behaviour involving
shocks so near the surface that the physical density there becomes
comparable to the atmosphere density. A typical value for the density
of the numerical atmosphere is $10^{-9}$ of the central neutron star
density, low in relative terms, but dense enough in absolute terms,
$10^9kg/m^3$, for a perfect fluid approximation to hold.

If one is mainly interested in the generation of gravitational waves
by the bulk motion of mass, the low-density, low-pressure exterior
regions can probably be safely neglected. However, photon and neutrino
emission are sensitive to temperature as well as mass, and therefore
could be dominated by shock heating at the surface. \new{Sound waves
  running down the density and pressure gradient towards the surface
have a tendency to shock, see \cite{GundlachPlease} for a
semi-quantitative criterion. In particular}, 
tidal forces in a
binary will raise waves in the interior of each star, which can form
shocks as they approach the surface, see \cite{gundlach-murphy}. Very
close to the surface, these shocks approach similarity solutions. This
means that they become singular, but at the same time that their
behaviour is universal, depending on the initial data only through a
single parameter.

As a step towards a correct numerical simulation of such phenomena in
astrophysical contexts, we construct the similarity solutions as the
solution of ordinary differential equations, and compute the related
similarity exponents. We then use the Clawpack software of \cite{claw.org}
to show numerically that the similarity solution is indeed an
attractor as the shock approaches the surface. To study the self-similarity
it is necessary to zoom in by many orders of magnitude on the origin (the
point where vacuum is reached) as the computation proceeds.  The Clawpack
software has been modified to repeatedly cut the domain in half and
redistribute the computational points over the new domain in order to
continue the calculation well into the asymptotic regime.  Similar
approaches have previously been used to study self-similarity numerically,
e.g. in the study of self-similar blow up in \cite{BergerKohn}.  Our
approach is described in Sec.~\ref{section:numerics}.

Throughout this paper, we assume that the fluid motion is
plane-parallel, and that the gravitational field is constant.  The
former is likely to be a good approximation for a sufficiently small
part of the surface, sufficiently close to the surface, but this
assumption will need to be tested in two or three-dimensional
simulations in future work. The approximation of constant
gravitational field as $g=GM/R^2$ is safe when combined with the
assumption of plane-parallel symmetry because the gravity of the outer
layers is neglible compared to the bulk mass $M$ of the star, and the
distance $R$ to the centre varies little. \new{As we are only
  considering a small region near the surface of star, general
  relativistic gravity can safely be approximated by Newtonian
  gravity, and the Lane-Emden and Tolman-Oppenheimer-Volkov solutions
  for spherical stars agree near the surface.}

We assume perfect fluid matter in the sense of local thermal
equilibrium, and consider two related equations of state.
\new{Throughout this paper, ``hot EOS'' will refer to
    the two-parameter equation of state $P(\rho,e)=(\Gamma-1)\rho e$
    often called the Gamma-law EOS, and ``cold EOS'' to the
    one-parameter (barotropic) equation of state
    $P(\rho)=K_0\rho^\Gamma$ often called the polytropic EOS. The
    former reduces to the latter if the entropy is constant. The
  latter does not allow for shock heating (and when we use it we do
  not solve the energy conservation law), and we consider it here
  mainly for comparison with the more physical hot EOS. In
  \cite{BaiottiGiacomazzoRezzolla2008}, these two equations, for
  $\Gamma=2$, are compared explicitly in neutron star binary mergers.}

In this paper, we consider only Newtonian fluid motion. We find that
this is self-consistent in the \new{cold} case. In the \new{hot} case,
the fluid velocity and sound speed calculated in Newtonian fluid
dynamics diverge as the shock approaches the surface, and so,
depending on the initial data and on how close to the surface the
perfect fluid approximation breaks down, one may have to go to special
relativity.

Secs.~\ref{section:equations} and \ref{section:shocks} review the
fluid equations and the Rankine-Hugoniot conditions at the shock.

Sec.~\ref{polytropicsection} generalises the similarity
solution of \cite{BarkerWhitham} from the
shallow water case $\Gamma=2$ to a \new{cold} 
polytropic gas with arbitrary $\Gamma$. We
add a discussion of the evolution after the shock reaches the surface.

Sec.~\ref{idealgassection} rederives the similarity solution of
\cite{Sakurai} for the non-relativistic motion \new{with the hot
  EOS}, and tabulates more cases. We add a discussion of the
equilibrium reached long after the shock has reached the surface.

In Sec.~\ref{section:numerics} we present details of the numerical
code, and numerical evidence that
generic smooth perturbations of the hydrostatic equilibrium are
attracted to the similarity solution as the surface is approached. 
Sec.~\ref{section:conclusions} presents our conclusions. 

In the Appendix, general mathematical facts about
power-law similarity solutions in one space and one time coordinate
are derived, and then applied \new{fluid
motion with our hot and cold EOS}. 
We discuss under what circumstances a conservation law in
the evolution equations gives rise to an integral of the motion in the ODE
system obtained under a similarity ansatz, and we show how the
similarity solutions with the \new{cold} equation of state relate to
the subclass of isentropic similarity solutions with the related
\new{hot} equation of state. 

In the following, $\simeq$ denotes
equality up to sub-leading terms, while $\sim$ denotes equality up to
sign, constant factors and sub-leading terms. In particular, our
convention is such that $t$ and $x$ are usually negative. To simplify
notation, we will take care of such signs when using the symbol
$\simeq$, but will neglect them when using $\sim$. The symbol $\equiv$
indicates definitions.


\section{The equations}
\label{section:equations}


\subsection{Conservation laws}
 

The 1D mass conservation and momentum conservation laws in
a constant gravitational field $g$ in the $-x$ direction are
\begin{eqnarray}
\rho_t+(v\rho)_x&=&0, \\
(v\rho)_t+(v^2\rho+P)_x&=&-g \rho.
\label{maineqns}
\end{eqnarray}
Here $\rho$ is the mass density, $v$ the velocity, and $P$ the pressure.
In the \new{cold} case, these equations are closed with the barotropic
equation of state
\begin{equation}
P(\rho)=K_0\rho^\Gamma, 
\end{equation}
Here $K_0$ and $\Gamma\equiv 1+1/n$ are constants with range $K_0>0$ and $n>0$
(and hence $\Gamma>1$). The $n=1$ cold case is equivalent to the
shallow water equations \new{(see \cite{rjl:fvmhp}) and the source term of
\eqn{maineqns} in this case corresponds to a linear beach with slope 1, so
that the similarity solution in this case shows the behavior of a bore on a
linear beach.}

By contrast, in the \new{hot} case the equation of state is
\begin{equation}
\label{gammalaw}
P(\rho,e)=(\Gamma-1)\rho e, 
\end{equation}
where the evolution of the internal energy per rest mass $e$ is given by
the total energy conservation law
\begin{equation}
\left[\rho\left(e+{v^2\over 2}\right)\right]_t+\left[\rho\left(e+{v^2\over
    2}\right)v+Pv\right]_x=-g\rho v.
\end{equation}


\subsection{Thermodynamic equations}


The first law of thermodynamics in the form
\begin{equation}
\label{firstlaw}
de=T\,ds-P\,d(\rho^{-1}),
\end{equation}
where $s$ is the entropy per rest mass and $T$ the temperature, can be
integrated over a curve of constant entropy in thermodynamical state
space, with $P$ given by (\ref{gammalaw}) to give
\begin{equation}
\label{egammalaw}
e(\rho,s)={K(s)\over \Gamma-1}\rho^{\Gamma-1},
\end{equation}
where $K(s)$ is some, \new{still unspecified,} function
of the entropy.  Hence we can write the \new{hot} equation of state in
the form
\begin{equation}
\label{Keos}
P(\rho,s)=K(s) \rho^{\Gamma}
\end{equation}
The \new{cold} equation of state can therefore be obtained as the
isentropic case $K=K_0={\rm const}$ of the \new{hot equation of state} with the
same polytropic index $\Gamma$ (or $n$). 

This form of the equation of
state can also be used to show that
\begin{equation}
\label{cPrho}
c^2\equiv \left. {\partial P\over\partial \rho}\right|_s
={\Gamma P\over\rho}
\end{equation}
for both the \new{hot and cold} case. (Physically, $c>0$
is the sound speed.) In the \new{hot} case, this can also be written as
\begin{equation}
\label{ce}
c^2
={\Gamma \over n}e. 
\end{equation}

\new{Making some additional independent assumption, such
  as the ideal gas law $P=RT\rho$, would determine the function $K(s)$, but
  for the remainder of this paper, we do not need to make such a choice.} 


\subsection{Riemann invariants}


In the \new{cold} case, the equations for smooth solutions can be
rewritten in the simpler form
\begin{equation}
\label{polyvc}
\left[\partial_t+(v\pm c)\partial_x\right](v\pm 2nc)+g=0. 
\end{equation}
For $g=0$, the equations therefore have a pair of Riemann invariants
$v\pm 2nc$, with characteristic speeds $v\pm c$ respectively. In the
case $g\ne 0$, the Riemann invariants still exist, and are $v+gt\pm
2nc$, with the same speeds. 

It is interesting to write the \new{hot} equations in almost Riemann
form as
\begin{eqnarray}
\left[\partial_t+v\partial_x\right]K&=&0, \\
\label{idealalmostRiem}
\left[\partial_t+(v\pm c)\partial_x\right](v\pm 2nc)+g&=&-{n\over
  \Gamma}c^2\partial_x \ln K,
\end{eqnarray}
where
\begin{equation}
K\equiv P\rho^{-\Gamma}
\end{equation}
is in physical terms a function of only the entropy per mass, and is
therefore conserved along particle trajectories. Note that in the
isentropic case the \new{hot evolution equations reduce to the cold ones}.


\subsection{Equilibrium solution}
\label{sec:equilibrium}


Under our assumptions, the equation of hydrostatic equilibrium is 
\begin{equation}
\label{hydrostatic}
P_x=-g\rho.
\end{equation}
In the \new{cold} case, the density in hydrostatic equilibrium, after
adjusting the origin of $x$, is given by
\begin{equation}
\rho(x)=\left(-{gx\over n\Gamma K_0}\right)^n
\end{equation}
which implies
\begin{equation}
\label{cequilibrium}
c(x)=\sqrt{-{gx\over n}}
\end{equation}
for $x<0$, where $x=0$ is the surface of the star. 

In the \new{hot EOS} case {\em in general}, the hydrostatic equilibrium
solution depends on the entropy stratification of the star. In the
astrophysics literature, a type of stratification is often considered where
\begin{equation}
p(x)=K_*\rho(x)^{\Gamma_*},
\end{equation}
where $K_*$ and $\Gamma_*$ are constants which describe the
equilibrium star. We define $\Gamma_*\equiv 1+1/n_*$. Note that
the equation of state is still given by (\ref{Keos}), with $K$ {\em not} in
general constant. Using the hydrostatic equilibrium condition
(\ref{hydrostatic}) and the expression (\ref{cPrho}) for the sound
speed, we find the generalised equilibrium solution
\begin{eqnarray}
\label{rhostratified}
\rho(x)&=&\left(-{gx\over n_*\Gamma_* K_*}\right)^{n_*}
\equiv C_\rho(-x)^{n_*}, \\
e(x) &=& -{gxn\over n_*\Gamma_*}, \\
\label{cstratequilibrium}
c(x)&=&\sqrt{-{gx\Gamma\over n_*\Gamma_*}}, \\
K(x)&=&K_*\left(-{gx\over n_*\Gamma_* K_*}\right)^{1-{n_*\over n}}.
\end{eqnarray}
(The last two expressions are redundant).  The case $n_*=n$ reduces to
the \new{cold} result above (with $K(x)=K_*$).  The range
$n<n_*<\infty$ gives a stratification that is stabilised by buoyancy
forces against (non-planar) vertical displacements. We will therefore
only consider $n_*\ge n$. The marginally stable limit $n_*=n$
corresponds to constant entropy, something that can be achieved by
convection, while $n_*=\infty$ can be interpreted as the isothermal
limit. We will not consider the latter, as the star then has no
surface at finite radius.


\section{Shock into fluid at rest}
\label{section:shocks}


\subsection{Mass and momentum conservation}


Let $v_s$, $\rho_s$, $P_s$ be the quantities just behind a shock
travelling in the positive $x$-direction (towards the surface at
$x=0$) with speed $\sigma$, with $v=0$ and $\rho_0$ and $P_0$ just in
front of the shock. Defining
\begin{equation}
\mu\equiv {\rho_0\over\rho_s},
\end{equation}
with the range $0<\mu<1$, the mass conservation law gives
\begin{equation}
\label{sigmalaw}
{v_s\over \sigma}=1-\mu.
\end{equation}
Using (\ref{cPrho}), and defining the dimensionless quantity
\begin{equation}
\lambda\equiv {c_0^2\over v_s^2}, 
\end{equation}
with range $0<\lambda<\infty$, the momentum
conservation law gives
\begin{equation}
\label{vslaw}
{\Gamma\over 1-\mu}=\left({P_s\over P_0}-1\right)\lambda.
\end{equation}


\subsection{\new{Cold equation of state}}


In the \new{cold} case 
\begin{equation}
{P_s\over P_0}=\mu^{-\Gamma}, 
\end{equation}
and so
\begin{equation}
\lambda={\Gamma \mu^\Gamma\over (1-\mu)(1-\mu^\Gamma)}.
\end{equation}
In the limit $\lambda\to 0$ we have $\mu\to 0$, with 
\begin{equation}
\label{epspoly}
\mu\simeq \left(\lambda\over\Gamma\right)^{1\over\Gamma}.
\end{equation}

In the opposite limit $\lambda\to\infty$, which will not be relevant
here, we have $\mu\simeq 1-\lambda^{-1/2}$.


\subsection{Energy conservation and the \new{hot equation of state} }


The energy conservation law gives
\begin{equation}
{P_s\over P_0}={{\Gamma(\Gamma-1)\over 2\lambda}-1
\over (1-\mu)\Gamma-1}
\end{equation}
and hence
\begin{eqnarray}
\mu&=&1+{\Gamma+1\over 4\lambda}-\sqrt{{1\over\lambda}+
\left({\Gamma+1\over 4\lambda}\right)^2}, \\
{P_s\over P_0}&=&1+\Gamma\left[{\Gamma+1\over 4\lambda}+\sqrt{{1\over\lambda}+
\left({\Gamma+1\over 4\lambda}\right)^2}\right],
\end{eqnarray}
where the sign of the square root is the correct one for the Lax shock.
In the limit $\lambda\to 0$, 
\begin{equation}
\label{epsideal}
\mu\simeq {\Gamma-1\over\Gamma+1}
\end{equation}
and 
\begin{equation}
{P_s\over P_0}\simeq {\Gamma(\Gamma+1)\over 2\lambda},
\end{equation}
which is equivalent to 
\begin{equation}
\label{eideal}
e_s\simeq {v_s^2\over 2}.
\end{equation}
This last result means that in the limit in which the internal energy
in front of the shock can be neglected, the internal energy behind the
shock is just the kinetic energy of the impact. 

In the opposite limit $\lambda\to\infty$, which will not be relevant
here, we have $\mu\simeq 1-\lambda^{-1/2}$. This is the same
limit as in the \new{cold} case, because for weak shocks shock heating
is negligible.


\section{The solution behind the shock \new{with the cold EOS}}
\label{polytropicsection}


\subsection{Matching conditions at the shock}


Tracing a Rieman invariant into the shock from behind, one
sees that with $c$ and $t$ finite, and the initial conditions finite,
$v_s$ must remain finite. Generically it will not be zero.  As the shock
approaches the surface of the star, where $c_0\to 0$, we 
have $\lambda\to 0$, and hence $\mu\to 0$. 

From (\ref{vslaw}) we then have $v_s\lesssim \sigma$. It is natural to
assume that $v_s$ has a limit $v_s\to v_*$ as the surface is
approached. This parameter $v_*$
is determined by the initial data. Therefore, near the surface we have
$v_s\simeq\sigma\simeq v_*$, and hence, after adjusting the origin of
$t$,
\begin{equation}
\label{xsapprox}
x_s\simeq v_* t.
\end{equation}
(Note that in our conventions $x_s<0$ and $t<0$.)
From (\ref{cequilibrium}), we have 
\begin{equation}
\label{c0approx}
c_0^2=-{gx_s\over n}\simeq -{gv_*t\over n},
\end{equation}
and hence 
\begin{equation}
\label{xsrough}
\lambda\simeq  -{gt\over nv_*}.
\end{equation}
From this with (\ref{epspoly}) we have
\begin{equation}
\mu\simeq  \left(-{t\over \tau}\right)^{1\over \Gamma},
\end{equation}
where we define
\begin{equation}
\label{TLdef}
\tau\equiv {n\Gamma v_*\over g}, \qquad \ell\equiv v_* \tau.
\end{equation}
Hence we find
\begin{equation}
\label{cspoly}
c_s=c_0\mu^{-{1\over 2n}}
\simeq \Gamma^{1/2}v_* \left(-{t\over \tau}\right)^{1\over
  2\Gamma}.
\end{equation}
To the same approximation (leading order in $\mu$), we can write 
(\ref{sigmalaw}) as
\begin{equation}
\label{vspoly}
\sigma -v_s \simeq \mu v_* \simeq v_* \left(-{t\over \tau}\right)^{1\over \Gamma}.
\end{equation}
Eqs.~(\ref{cspoly}) and (\ref{vspoly}) provide
boundary conditions at the shock for the smooth solution behind it.
Our derivation of Eq.~(\ref{cspoly}) generalises the ``Whitham
approximation'' of \cite{KellerLevineWhitham} from the shallow water
case $n=1$ to general $n$.

We see that $K_0$ drops out of the result when expressed in terms of $c$
rather than $\rho$, but both $v_*$ and $g$ are relevant. (Note that
$g$ comes in not through its effect on the fluid behind the shock,
which we neglect, but from the assumption that the fluid in front of
the shock is in hydrostatic equilibrium.)


\subsection{Similarity solution}


The exact solution behind the shock is likely to be attracted to a
similarity solution, with the initial data remembered only through the
parameter $v_*$. To remove $v_*$ as far as possible, we go to a frame
moving with constant velocity $v_*$, and define the new variables
\begin{equation}
\label{uxidef}
u\equiv v-v_*, \qquad \xi\equiv x-v_* t
\end{equation} 
The PDEs that apply behind the shock are Galileo-invariant, and so
(\ref{polyvc})
becomes
\begin{equation}
\label{polyuc}
\left[\partial_t+(u\pm c)\partial_\xi\right](u\pm 2nc)+g=0,
\end{equation}
but in the shock conditions we must take into account that the
unperturbed fluid has $v=0$. Therefore we also define
\begin{equation}
s\equiv {d\xi_s\over dt}= \sigma -v_*, 
\end{equation}
and note
\begin{equation}
\label{sus}
s-u_s=\sigma-v_s.
\end{equation}
It is clear that in the limit $t\to 0$, we can neglect the
gravitational force compared to pressure forces, and so we look for a
solution with $g=0$.

We can write the similarity solution as
\begin{eqnarray}
\label{csim}
c(x,t)&=&{\xi\over t}\hat c(y) \\
\label{usim}
u(x,t)&=&{\xi\over t}\hat u(y),
\end{eqnarray}
where
\begin{equation}
\label{ydefxi}
y\equiv -{\xi\over \ell} \left(-{t\over \tau}\right)^{-\beta},
\end{equation}
with $\beta$ still undetermined. (See Appendix~\ref{appendix:polytropic} for a
derivation.) 

Above, we have shown that the matching conditions at the shock
require
\begin{eqnarray}
\label{eq1}
c_s &\sim& t^{1\over 2\Gamma},\\
\label{eq2bis}
s-u_s &\sim& t^{1\over\Gamma} \ll c_s
\end{eqnarray}
as $t\to 0_-$, and so 
\begin{equation}
\label{vspolyapprox}
s \simeq u_s. 
\end{equation}
From the last equation, comparing with (\ref{usim}), we have $y_s\sim
{\rm const}$ to leading order. The matching
condition (\ref{vspolyapprox}) gives 
\begin{equation}
\label{polytropiccond1}
\hat u(y_s)=\beta.
\end{equation}
Note that the more
accurate matching condition (\ref{vspoly}) could only be imposed at a
higher order in $\lambda$ as $\lambda\to 0$.
The matching condition (\ref{cspoly}) then gives
\begin{eqnarray}
\label{polytropiccond2}
y_s \hat c(y_s)&=&\Gamma^{1/2},\\
\label{mybeta}
\beta&=&{2+3n\over 2+2n}={2\Gamma+1\over 2\Gamma},
\end{eqnarray}
for the overall factor and exponent, respectively.  

The relevant similarity solution, for arbitrary $\beta$, can be found
in closed form under the assumption that $u+2nc=0$, which means that
the Riemann invariant running into the shock from behind vanishes (see
Appendix~\ref{appendix:polytropic}), or physically, that the solution
  behind the shock is dominated by a wave travelling backwards from
  the shock. This solution can be found in implicit form as
\begin{eqnarray}
\label{exactsolutionc}\hat c(y)^\beta \left(\hat c(y)+{1\over
  2n+1}\right)^{1-\beta}&=&{A\over y},\\
\label{exactsolutionu}
\hat u(y)+2n\hat c(y)&=& 0.
\end{eqnarray}
The matching conditions (\ref{polytropiccond1}) and
(\ref{polytropiccond2}) then fix
\begin{eqnarray}
A&=&\sqrt{\Gamma}\left({2+3n+2n^2\over 2+7n+6n^2}\right)^{-{n\over
      2+2n}}, \\
y_s &=&-{4\sqrt{n}(1+n)^{3/2}\over 2+3n}.
\end{eqnarray}

We can write
\begin{equation}
c(x,t)=v_*\left(-{t\over \tau}\right)^{\beta-1}\bar c(y), 
\quad \bar c(y)\equiv y \hat c(y),
\end{equation}
and similarly for $u(x,t)$. Hence $c(x,t)$ is a regular function of
$x$ at $t<0$ if and only if $\bar c(y)$ is a regular function of
$y$. From (\ref{exactsolutionc}) the latter is the case for the range
$y_{\rm min}<y<\infty$ for some $y_{\rm min}<0$. 

\begin{figure}
\hfil\includegraphics[width=7cm]{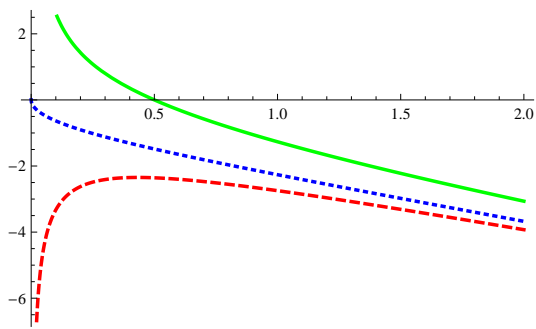} \hfil
\caption{ 
\label{fig:polytropicyrange}
\new{Cold EOS}: Plots of (from below) $y_{\rm min}$ (lower boundary of the domain of the
similarity solution, red, dashed), $y_s$ (shock location, blue, dashed) and $y_c$
(sonic point, green, solid) against the polytropic index $n$.}
\end{figure}

\begin{figure}
\hfil\includegraphics[width=7cm]{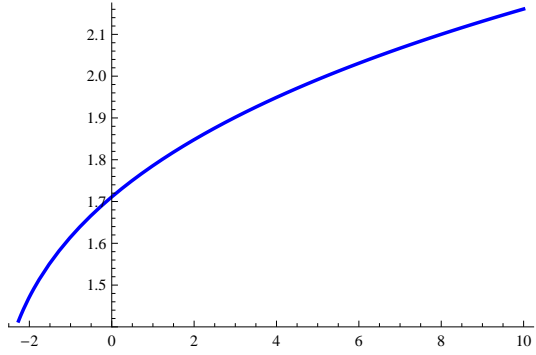} \hfil
\caption{ 
\label{fig:ychatplotn1}
\new{Cold EOS}: Plot of $\bar c(y)$ against $y$ over the range $y_s\le
y\le 10$, for $n=1$. This plot gives the instantaneous
profile of the sound speed, compare Eqs.~(\ref{csim}) and
(\ref{ydefxi}).}
\end{figure}

As discussed in Appendix~\ref{appendix:polytropic}, the sonic point
$y=y_c$ is a line of constant $y$ which is also a fluid
characteristic. It is given by the condition $\hat
c=-\beta/(2n-1)$. Fig.~\ref{fig:polytropicyrange} proves graphically
that $y_{\rm min}<y_s<y_c<\infty$ for all $n>0$. (The expressions for
$y_{\rm min}$ and $y_c$ are known in closed form but are long.)
Therefore, for $t<0$ the similarity solution exists between the shock
location $x=x_s$ and $x=-\infty$, passing through a sonic point
$x=x_c$ closely behind the shock. Fig.~\ref{fig:ychatplotn1} shows
$\bar c(y)$ against $y$ for $n=1$.

We can also write
\begin{equation}
c(x,t)= v_*\left(-{\xi\over \ell}\right)^\gamma \tilde c(y), 
\quad \tilde c(y)\equiv y^{1\over\beta}\hat c(y),
\end{equation}
and similarly for $u(x,t)$, where
\begin{equation}
\gamma\equiv1-{1\over\beta}={n\over 2+3n}={1\over 2\Gamma+1}.
\end{equation}
The limit 
\begin{equation}
\tilde c(\infty)=A^{1/\beta}(2n+1)^{-\gamma}
\end{equation}
exists, and so $u(x,t)$ and $c(x,t)$ are regular as $t\to 0_-$, and
become simply
\begin{equation}
\label{yinftylimit}
c(x,0)=v_*\tilde c(\infty)\left(-{x\over \ell}\right)^\gamma, \quad
u(x,0)=-2nc(x,0).
\end{equation}

\begin{table}
\begin{centering}
\begin{tabular}{c|cccccc}
$\Gamma$ & $\beta$ & $\gamma$ & $A$ & $y_s$ & $y_c$ & $\tilde c(\infty)$ \\
\hline
 1.1 & 1.45455 & 0.3125 & 1.70076 &  -14.276 & 1.70076 & 0.556382 \\
 1.2 & 1.41667 & 0.294118 & 1.68007 &  -7.45382 & 1.68007 & 0.712463 \\
 1.3 & 1.38462 & 0.277778 & 1.66785 &  -5.08952 & 1.66785 & 0.821717 \\
 4/3 & 11/8 & 3/11 & 1.66534 & -4.60059 & 1.66534 & 0.85234 \\
 1.4 & 1.35714 & 0.263158 & 1.66236 &  -3.84899 & 1.66236 & 0.907535 \\
 1.5 & 1.33333 & 0.25 & 1.66223 &  -3.06428 & 1.66223 & 0.978987 \\
 1.6 & 1.3125 & 0.238095 & 1.66638 & -2.51179 & 1.66638 & 1.04075 \\
 5/3 & 13/10 & 3/13 & 1.67109 &  -2.22244 & 1.67109 & 1.07795 \\
 1.7 & 1.29412 & 0.227273 & 1.67394 &  -2.09497 & 1.67394 & 1.09559 \\
 1.8 & 1.27778 & 0.217391 & 1.68424 & -1.76503 & 1.68424 & 1.14528 \\
 1.9 & 1.26316 & 0.208333 & 1.69676 & -1.49452 & 1.69676 & 1.19101 \\
 2 & 5/4 & 1/5 & 1.71105 & -1.26671 & 1.71105 & 1.23363 \\
\end{tabular}
\caption{
\label{table:polytropic}
\new{Cold EOS}: Table of approximate numerical values (the closed form
expressions are given in the text) of $\beta$,
$\gamma$, $A$, $y_s$, $y_c$ and $\tilde c(\infty)$ against the polytropic index
$\Gamma$. }
\end{centering}
\end{table}


\subsection{Evolution after the shock has reached the surface}


The data (\ref{yinftylimit}) at $t=0$ are smooth, and their evolution
to $t>0$ remains smooth.  
We were able to neglect gravity in the
approach of the shock to the surface, but we must restore it in the
subsequent smooth motion which occurs on a much longer timescale. This
can be done exactly by going to a freely falling frame by replacing
$u$ and $\xi$ with
\begin{eqnarray}
\zeta &\equiv& x + v_*t-{1\over 2} gt^2, \\
w &\equiv& v-v_*+gt.
\end{eqnarray}
The equations become 
\begin{equation}
\label{polywc}
\left[\partial_t+(w\pm c)\partial_\zeta\right](w\pm 2nc)=0. 
\end{equation}
Clearly $w+2nc=0$ at $t=0$ and hence for $t>0$, and so the solution
for $t>0$ is a simple wave characterised by
\begin{equation}
c_t-(2n+1)cc_\zeta=0.
\end{equation}
The method of characteristics gives $c(\zeta,t)$ implicitly as
\begin{equation}
c=f\left[\zeta+(2n+1)ct\right],
\end{equation}
where $f(x)=c(x,0)$ given in (\ref{yinftylimit}).


\section{The solution behind the shock \new{with the hot EOS}}
\label{idealgassection}


\subsection{Matching conditions at the shock}


\new{Our numerical experiments show}
that the fluid velocity $v_s$ behind the shock does
not remain constant but blows up as the shock approaches the
surface. This implies that once again we are in the regime $\lambda\to
0$. From (\ref{sigmalaw}) with (\ref{epsideal}) we obtain
\begin{equation}
\sigma\simeq {\Gamma+1\over 2} v_s.
\end{equation}
With (\ref{ce}), (\ref{eideal}) gives us
\begin{equation}
\label{csideal}
c_s\simeq \sqrt{\Gamma \over 2n} v_s.
\end{equation}
We need a third matching condition for the
density. From (\ref{epsideal}), we have
\begin{equation}
\rho_s\simeq{\Gamma+1\over\Gamma-1}\rho_0.
\end{equation}
Here the equilibrium density in front of the shock $\rho_0$ is given
by Eq.~(\ref{rhostratified}).


\subsection{Similarity solution}
\label{section:idealgassimilarity}


The similarity ansatz for the velocity and
sound speed is similar to (\ref{csim}-\ref{ydefxi}), namely 
\begin{eqnarray}
\label{csimideal}
c(x,t)&=&{x\over t}\hat c(y), \\
v(x,t)&=&{x\over t}\hat u(y), \\
\rho(x,t)&=& C_\rho(-x)^{n_*} \tilde \rho(y),
\end{eqnarray}
where
\begin{equation}
\label{ydefideal}
y\equiv C_y(-x)(-t)^{-\beta},
\end{equation}
where $C_y$ is a parameter (of dimension $L^{-1}T^\beta$) that
survives from the initial data into the self-similar regime.

The relevant similarity solution begins at $y=\infty$ (corresponding
to $x=-\infty$ at constant $t<0$), goes through a regular sonic point
$y=y_c$, and ends at the shock at $y=y_s$. As the similarity equations
are autonomous in $\eta=\ln y$, we can assume that the sonic point is
at $y=1$. 

We expand the solution through the regular sonic point
as $\hat u=\hat u_0 +\hat u_1 \eta + O(\eta^2)$ and $\hat c=\hat c_0
+\hat c_1 \eta + O(\eta^2)$. For given $n_*$, $\beta$ and $n$, we
obtain a unique value of $(\hat u_0,\hat c_0)$, but two values of
$(\hat u_1,\hat c_1)$. These correspond to two smooth solution curves
going through the same regular sonic point. The correct solution is
easily identified by plotting.

With the shock at $\eta=\eta_s$, the matching conditions at the shock give
\begin{eqnarray}
\hat u(\eta_s)&=&{2\beta\over\Gamma+1}, \\
\hat c(\eta_s)&=&\sqrt{\Gamma\over 2n}\hat u(\eta_s), \\
\tilde\rho(\eta_s)&=& {\Gamma+1\over \Gamma-1}.
\end{eqnarray}
The requirement that for given $n_*$ and $n$ the solution curve
$(\hat u(\eta),\hat c(\eta))$ goes smoothly through the sonic point and
ends at the shock then determines $\beta$ and $\eta_s<0$ through a boundary
value problem. 

We solve this boundary value problem numerically by
shooting from the sonic point, or rather from a small negative value
of $\eta$ reached by the power-law expansion around $\eta=0$, to a
trial value of $\eta_s$, using a trial value of $\beta$. An accurate
first guess for $\eta_s$ and $\beta$ is obtained by approximating the
solution curve from the sonic point to the shock as a linear function
of $\eta$ using $\hat u_1$ and $\hat c_1$. 

As in the \new{cold} case, we can write
\begin{eqnarray}
\label{cbarideal}
c(x,t) &=& C_y^{-1}(-t)^{\beta-1} \bar c(y) \\
&=& C_y^{-{1\over\beta}}(-x)^\gamma \tilde c(y), 
\end{eqnarray}
and similarly for $u(x,t)$. The limits $\tilde c(\infty)$ and $\tilde
u(\infty)$ exist and are tabulated in Table~\ref{table:isentropic}. 
In contrast to the
\new{cold} case, $\gamma<0$, and so $c$ and $v$ behind the shock
diverge as $x\to 0_-$. Similarly, the instantaneous values of $c$ and
$v$ just behind the shock diverge because $\beta-1<0$. In physical terms
this can be thought of as a whiplash effect, where a finite amount of
kinetic and internal energy is injected into ever less mass.

We can write the instantaneous density profile at $t<0$ as
\begin{equation}
\label{rhobarideal}
\rho(x,t)=C_\rho C_y^{-{n_*}} (-t)^{n_*\beta} \bar \rho(y),
\quad \bar\rho(y)\equiv y^{n_*} \tilde\rho(y).
\end{equation}
Finally, the limit $\tilde\rho(\infty)$ exists and is tabulated
below. Therefore the density immediately behind the shock, and the
density everywhere as $t\to 0_-$ remain finite, as one would expect
from the fact that the shock conditions show a finite compression
ratio. The density profile at $t=0$ is just the one in the rest state
multiplied by $\tilde\rho(\infty)$. This constant can be found
explicitly in terms of other known constants by evaluating the
integral (\ref{integral}) of the similarity equations at $y=y_s$ and
as $y\to\infty$.

As a test of our calculation, Table~\ref{table:sakurai} reproduces
Table~1 of \cite{Sakurai}. Focussing then on the case of
an isentropic rest state, Table~\ref{table:isentropic} gives key
parameters of the similarity solutions for selected values of
$\Gamma$. 

\begin{table}
\begin{centering}
\begin{tabular}{c|ccc}
$\Gamma\backslash n_*$ & 2 & 1 & 1/2 \\
\hline
5/3 & 0.435628 & 0.22336 & 0.1141 \\
7/5 & 0.3934 & 0.202151 & 0.103519 \\
6/5 & 0.330985 & 0.170418 & 0.0874924 \\
\end{tabular}
\caption{
\label{table:sakurai}
Nonrelativistic \new{hot EOS}: Recreation of Table~1 of \cite{Sakurai}:
  $1/\beta$ tabulated against $\Gamma$ (down) and $n_*$ (right).}
\end{centering}
\end{table}

\begin{table}
\begin{centering}
\begin{tabular}{c|ccc}
$n\backslash n_*$ & 3 & 3/2 & 1 \\
\hline
  3 & 0.642161 & &  \\
  3/2 & 0.608205 & 0.751791 &  \\
 1 & 0.592366 & 0.739586 & 0.807808 \\
\end{tabular}
\caption{
\label{table:stratified}
Nonrelativistic \new{hot EOS}: Table of $\beta$ against $n$ and
$n_*$. We only consider stable stratifications where $n_*\ge n$.} 
\end{centering}
\end{table}
\begin{table}
\begin{centering}
\begin{tabular}{c|ccccc}
$\Gamma$ & $\beta$ & $\eta_s$ & $\tilde u(\infty)$ &
  $\tilde c(\infty)$ & $\tilde\rho(\infty)$ \\
\hline
 1.1 &  0.436716 & -0.0211101 & 0.127454 & 0.0190103 & 765.341 \\
 1.2 &  0.555056 & -0.0468527 & 0.303072 & 0.0664771 & 2.01223 \\
 1.3 &  0.624385 & -0.0722611 & 0.356576 & 0.101789 & 1.11067 \\
 4/3 &  0.642089 & -0.0802736 & 0.368456 & 0.113245 & 0.994484 \\
 1.4 &  0.672222 & -0.0954616 & 0.385789 & 0.135492 & 0.864565 \\
 1.5 &  0.707991 & -0.116087 & 0.400758 & 0.167296 & 0.808397 \\
 1.6 &  0.736085 & -0.134239 & 0.410263 & 0.198712 & 0.743478 \\
 5/3 &  0.751778 & -0.14509 & 0.413301 & 0.218764 & 0.363108 \\
 1.7 &  0.758898 & -0.150172 & 0.414418 & 0.228711 & 0.673344 \\
 1.8 &  0.777879 & -0.164173 & 0.414795 & 0.257156 & 0.655824 \\
 1.9 &  0.793966 & -0.176514 & 0.412426 & 0.283948 & 0.63714  \\
 2   & 0.807803 & -0.187436 & 0.408344 & 0.309271 & 0.62677 \\
\end{tabular}
\caption{
\label{table:isentropic}
Nonrelativistic \new{hot EOS}: Table of the parameters $\beta$, $\eta_s$
and asymptotic values at $y=\infty$ of $\tilde u$, $\tilde c$ and
$\tilde \rho$, against the polytropic index $n$, for the isentropic
rest state case $n_*=n$. $\Gamma>2$ is not considered because the
sound speed at ultrarelativistic temperatures is greater than the
speed of light. $\Gamma=1$ is unphysical because the speed of sound is
zero. We also have difficulty reaching $\Gamma=1$ when numerically
solving the boundary value problem for the similarity solution.}
\end{centering}
\end{table}

\begin{figure}
\hfil\includegraphics[width=7cm]{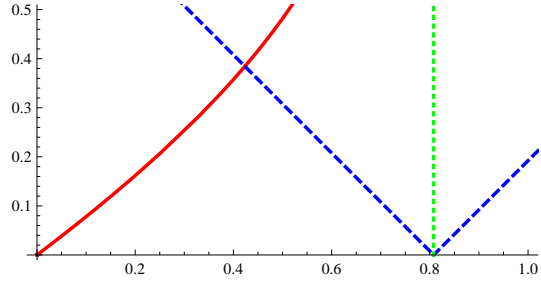} 
\caption{ 
\label{fig:idealgasdynsys}
Nonrelativistic \new{hot EOS}: Plot of the similarity solution with
$n=n_*=1$ \new{in the $\hat u\hat c$ plane,
    illustrating the dynamical systems ideas discussed in
    Sec.~\ref{section:idealgassimilarity}}. $\hat u$ is right, and
$\hat c$ is up. The vertical straight line is the flow line and the
diagonal straight line are the sonic lines (see
Apppendix~\ref{idealgassimilarity}). The endpoint at $\hat u=\hat c=0$
corresponds to $y=\infty$, while the other endpoint corresponds to the
shock $y=y_s$, with the sonic point at $y_c=1$ by construction.}
\end{figure}

\begin{figure}
\hfil\includegraphics[width=7cm]{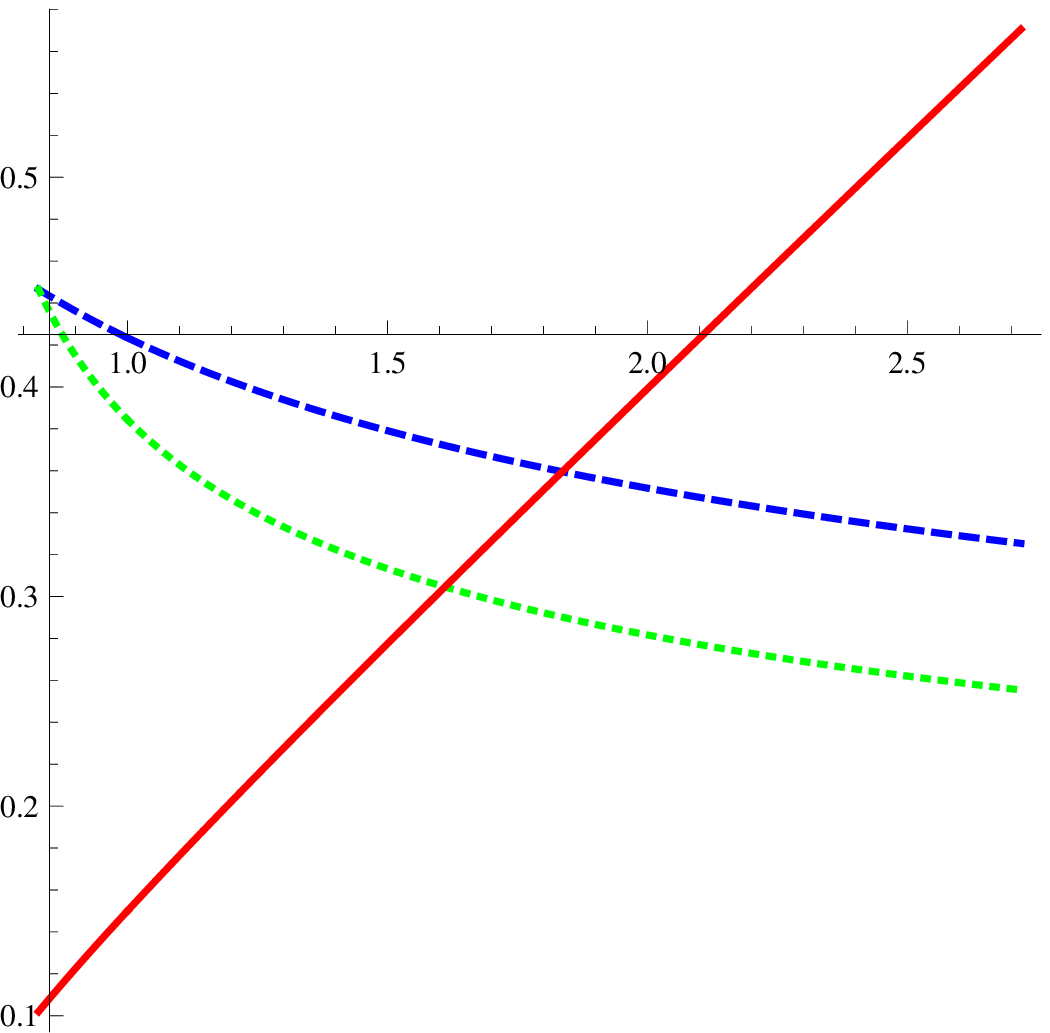} 
\caption{ 
\label{fig:idealgasprofiles}
Nonrelativistic \new{hot EOS}: Plots of (from above) $\bar u(y)$
(blue, dashed), $\bar c(y)$ (green, dotted) and $\bar\rho/20$ (red,
solid) for the similarity solution with $n=n_*=1$ for the range
$y_s<y<e^1$. The sonic point is at $y_c=1$ by definition. These curves
give the instantaneous profiles of $u$, $c$ and $\rho$ against $x$ at
finite $t<0$, compare Eqs.~(\ref{ydefideal}), (\ref{cbarideal}) (and
similarly for $\bar u$) and (\ref{rhobarideal}).  }
\end{figure}


\subsection{New hydrostatic equilibrium after the shock}


Consider now an arbitrary instantaneous state in which density and
internal energy follow approximate power laws,
\begin{eqnarray}
\label{adef}
\rho &\sim& x^a, \\
\label{bdef}
c_s^2 \sim T \sim e &\sim& x^b.
\end{eqnarray}
Note that at $t=0$ in the shock heating process we have discussed,
$b=2\gamma$ and $a=n_*$. By contrast, Eq.~(\ref{cstratequilibrium}) tells us
that in hydrostatic equilibrium $b=1$.

From (\ref{adef}) and (\ref{bdef}) we have
\begin{equation}
\label{Kxpower}
K\sim x^{b-{a\over n}}.
\end{equation}
We define the total mass per area (measured from the surface inwards)
as 
\begin{equation}
m\equiv \int_x^0\rho\,dx \sim x^{a+1}
\end{equation}
and hence
\begin{equation}
K\sim m^{b-{a\over n}\over a+1}.
\end{equation}

Let us now assume that the fluid reaches a new hydrostatic equilibrium
starting from the state (\ref{adef},\ref{bdef}). This new equilibrium
has a density distribution with power $\bar a$ and by definition has
$\bar b=1$. If heat conduction, shock heating and other
entropy-generating processes can be neglected in the approach to
equilibrium, and if the fluid motion remains plane (spherically)
symmetric, then the function $K=K(m)$ is unchanged, because $m$ is a
Lagrangian coordinate for the plane-symmetric motion. In particular,
we must have
\begin{equation}
{b-{a\over n}\over a+1}={1-{\bar a\over n}\over \bar a+1},
\end{equation}
which solves to 
\begin{equation} 
\bar a={(n+1)a+n(1-b)\over 1+bn}.
\end{equation}


\section{Numerical experiments}
\label{section:numerics}


\subsection{Numerical method}

To confirm the validity of the similarity solutions presented here, we have
done extensive numerical simulation using a high-resolution finite volume
method designed to accurately capture shock waves.  We use the open-source
Clawpack software of \cite{claw.org}.
The codes used to produce the figures in this section are
available at \cite{cg-rjl:shockvacuum10.web}
along with more figures and animations of the results over time.

This software requires
a ``Riemann solver'' as the basic building
block. For a homogeneous conservation law with no source term, the Riemann
solver takes cell averages in two neighboring grid cells and
determines a set of waves propagating away from the cell interface in the
solution to the Riemann problem (the conservation law with piecewise constant
intial data).  A simple update of the cell averages based on the distance
these waves propagate into the cells gives the classic Godunov method, a
robust but only first-order accurate numerical method.  Second order
correction terms can be defined in terms of these waves and then limiters
are applied to these terms in order to avoid non-physical oscillations 
in the solution.  This is crucial near discontinuities
for problems involving shock waves, particularly when near a vacuum state as
in the present problem.  
Complete details of the algorithms implemented in Clawpack can be found in
\cite{rjl:fvmhp}, \cite{rjl:wpalg}.  
Several other books also discuss shock-capturing finite volume methods based
on Riemann solvers, such as \cite{toro} and \cite{trangenstein:hyper}.
The development of high-resolution shock-capturing
methods has a long history, and many references can be found in the books
cited above.

The discussion here will focus on
several modifications that have been made to the standard
software in order to handle the demands of the present problem.
We use the f-wave formulation of the the
wave-propagation algorithm originally proposed in \cite{db-rjl-sm-jr:vcflux}
in order to obtain a ``well-balanced'' method that preserves the steady
state equilibrium solution in the stationary atmosphere ahead of
the shock wave, following the approach in \cite{rjl:wbfwave10}.  
Maintaining this steady state requires that the pressure
gradient in  \eqn{maineqns} balances the source term $-g\rho$.  
In the f-wave formulation, the flux difference between adjacent cells is
modified by the source terms and the Riemann solver applied to this modified
flux difference to obtain f-waves that are used in place of the classic
waves of the Riemann solution.  If the source terms are appropriately
discretized at each interface, the method will be well balanced in the sense
that initial data that are
in equilibrium will lead to zero-strength f-waves in each
Riemann solution and hence no modification to the solution.  Moreover, small
perturbations to an equilibrium solution will result in small amplitude
f-waves.  The limiters are applied to these f-waves and much more accurate
solutions can be obtained for certain quasi-steady problems with this
approach than with techniques such as fractional step methods (in which one
alternates between solving the homogeneous conservation law and applying the
source terms).  Discretization of the source term $-g\rho$ requires defining an
appropriate average of the
densities from the two cells bordering each interface and we use the
approach suggested in \cite{rjl:wbfwave10} to achieve a well-balanced method
for the atmosphere at rest.

The Riemann solver for the f-wave method must then take the flux difference
between two cells, as modified by the source term, and split this vector
into waves propagating to the left and right that are used to update
neighboring cell averages.  This is typically done by splitting the vector
into eigenvectors of some approximate Jacobian matrix such as the one
produced by a Roe average of the neighboring states.   However, near vacuum
state the standard approach can lead to negative pressures or densities and 
a breakdown of the method.  We use a variant of the Siliciu relaxation
solver described in Sections 2.4.4--6 of \cite{bouchut}, which is 
related to the HLLE solver.  This solver has been modified to work with the
f-wave formulation in recent work with \cite{murphy:pc}.

To see the self-similar structure that eventually develops, we need to
zoom in by many orders of magnitude in the vicinity of $x=0$ at the
times just before the shock reaches the origin.  For general initial
data specified at some time $t_0$ it is impossible to determine {\em a
  priori} the exact final time $t_f$ when the shock will reach $x=0$,
and it varies depending on the mesh spacing and the choice of
computational parameters such as the limiter used to maintain
stability. 
\new{For the computations presented here we have used the monotonized
centered (MC) limiter (see \cite{rjl:fvmhp}), which is generally a good
limiter for
nonlinear systems that gives sharper results than the minmod limiter but
greater stability than superbee, for example.  We have also tried minmod
and obtain similar results.}
Note that the parameter $t$ used elsewhere in the paper for
the similarity solution corresponds to $t=t_c-t_f$ where $t_c$ is the
time variable in the computation. \new{We start the computation with
  $t_c=0$ but reset to $t_c=0$ at some point as we approach the final
  time in order to retain more significant figures in $t$.}

We compute on a domain $\xmin\leq x\leq 0$ which is divided into $M$ finite
volume cells. Initially $\xmin=-12$, but we repeatedly cut the domain in
half by halving both $\xmin$ and $\Dx$ in order to zoom in on the origin.  
In each time step we estimate the current shock location $x_s$ 
\new{as the location of the largest change in density} and we
perform this regridding as soon as $x_s>0.1\xmin$, i.e., when the shock is
90\% of the way to the boundary.  Regridding is performed by taking the
solution on the right half of the domain and doubling the resolution, so
that $M$ remains the same while $\xmin$ and $\Dx$ are halved.  In each grid
cell $j=M/2+1,~\ldots,~M$ we estimate a slope for each component of the
solution 
\new{by applying a limiter to the one-sided slopes to the left
and right, and use this to construct a linear function passing through the
cell average with this slope (essentially identical results are obtained
with either the minmod or superbee limiters). }
We sample the resulting linear function at the midpoint of
the left and right halves of the cell.  These two values replace the values
stored in cells $i$ and $i+1$, where $i=2(j-M/2) - 1$.
This procedure is conservative since the two new values have the same mean
value as the previous single value.  It is second order accurate where the
solution is smooth, but it does introduce some artifacts at the shock.  The
new interpolated values do not lie exactly on the numerical shock Hugoniot
for a propagating shock and hence small oscillations generally emerge from
the shock at the regridding times that propagate into the flow behind the
shock.  These are quite minor and do not hamper our ability to see the
expected self-similar behavior.

The time step $\Dt$ is chosen automatically by the Clawpack software, based
on a user-supplied target Courant number, which we take to be 0.8.  The
Courant number is computed in each time step as
\eql{cfl}
\CFL = \frac \Dt \Dx \max_{i,p} |s_i^p|
\end{equation}
where $s_i^p$ is the wave speed of the $p$th wave for the Riemann problem
between cells $i-1$ and $i$.  The time step $\Dt$ for the next step is
chosen by setting $\CFL$ in \eqn{cfl} to the target Courant number and
solving for $\Dt$.  (The actual Courant number in the next step may 
differ from the target since the nonlinear wave speeds $s_i^p$ are
recomputed in each step.  If the actual Courant number is larger than the
stability limit of 1 then the step is re-taken with a smaller $\Dt$, which
at this point can be calculated to hit the target Courant number
exactly since the waves
speeds are now known.)  Using this mechanism, the time step automatically
scales with $\Dx$ as the computational domain is subdivided.  We output
results every $\Nout$ time steps (some fixed number depending on $M$) to
obtain snapshots of the solution as it gets further into the asymptotic
regime.  The space-time numerical grids used resemble a set of nested boxes 
all sharing the top right corner at $x=0,~t_c=t_f$.
Note that with this strategy the shock never reaches $x=0$ and we typically
halt the computation when $\xmin \approx 10^{-10}$ (after about 35 domain
halvings). 

From the final frames of the solution we can estimate $t_f$, the time when
the shock would hit $x=0$.  As mentioned above, to obtain enough significant
figures in $t_f-t_c$ we reset $t_c$ to 0 at some point in the computation,
typically when $\xmin \approx -10^{-6}$.

For boundary conditions at $x=\xmin$ we use the zero-order extrapolation
conditions that are often used in Clawpack to 
avoid non-physical reflections at computational boundaries.  
The values in the first
interior grid cell are simply copied to ghost cells adjacent to the cell
before each time step as described in Chapter 5 of \cite{rjl:fvmhp}.
This is obviously unphysical in that it does not represent
the overall solution that we have cut off. However, this has no influence
on the portion of the solution we are interested in.
After regridding, the shock has been shifted from 
$x\approx 0.1\xmin$ to $x\approx 0.2\xmin$ after $\xmin$ is halved.  
Using an explicit method 
with Courant number less than 1, information can propagate no more than 1
grid cell per time step.  Hence in the time between regriddings any effect
of boundary conditions at $x=\xmin$ can contaminate at most 10\% of the
cells before the next regridding takes place, at which point we throw away
the left half of the cells.  So these boundary conditions can never affect the
solution near the shock wave.

The boundary condition imposed at $x=0$ is immaterial for this work since we
never allow the shock to reach the boundary.  We need only insure that the
equilibrium solution is not disturbed adjacent to the boundary.  We use a
well-balanced method as described above that maintains the steady state
between regridding times.  Unfortunately, this is disturbed by regridding
since reconstructing and sampling a piecewise linear function as described
above gives new cell values that are not in exact equilibrium.  So in the
undisturbed region ahead of the shock we reset the solution by evaluting the
exact form of the known equilibrium solution on the new finer grid
each time we regrid.


\subsection{\new{Cold} equation of state}

\begin{figure}
\hfil\includegraphics[width=6cm]{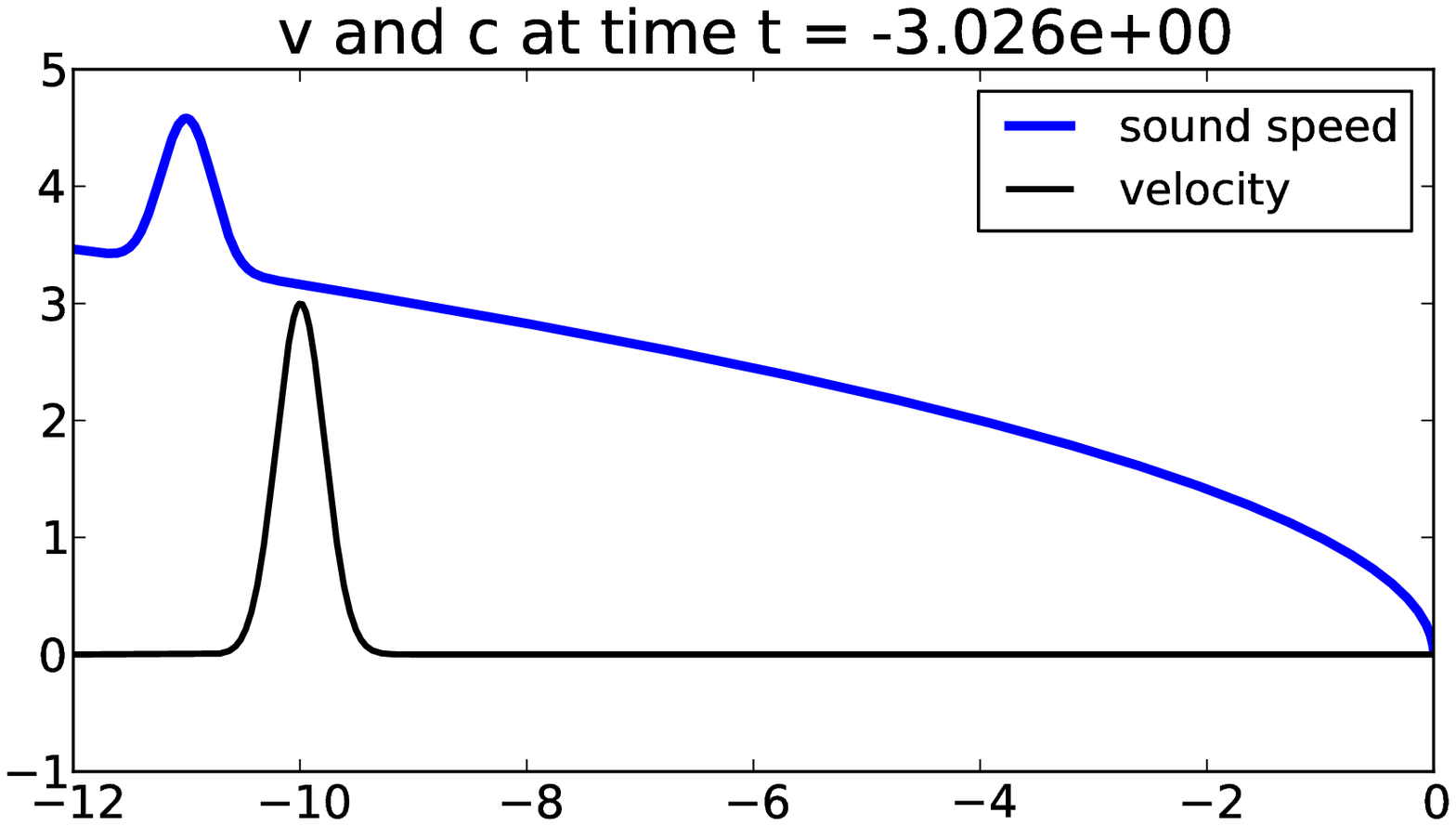} 
\hfil\includegraphics[width=6cm]{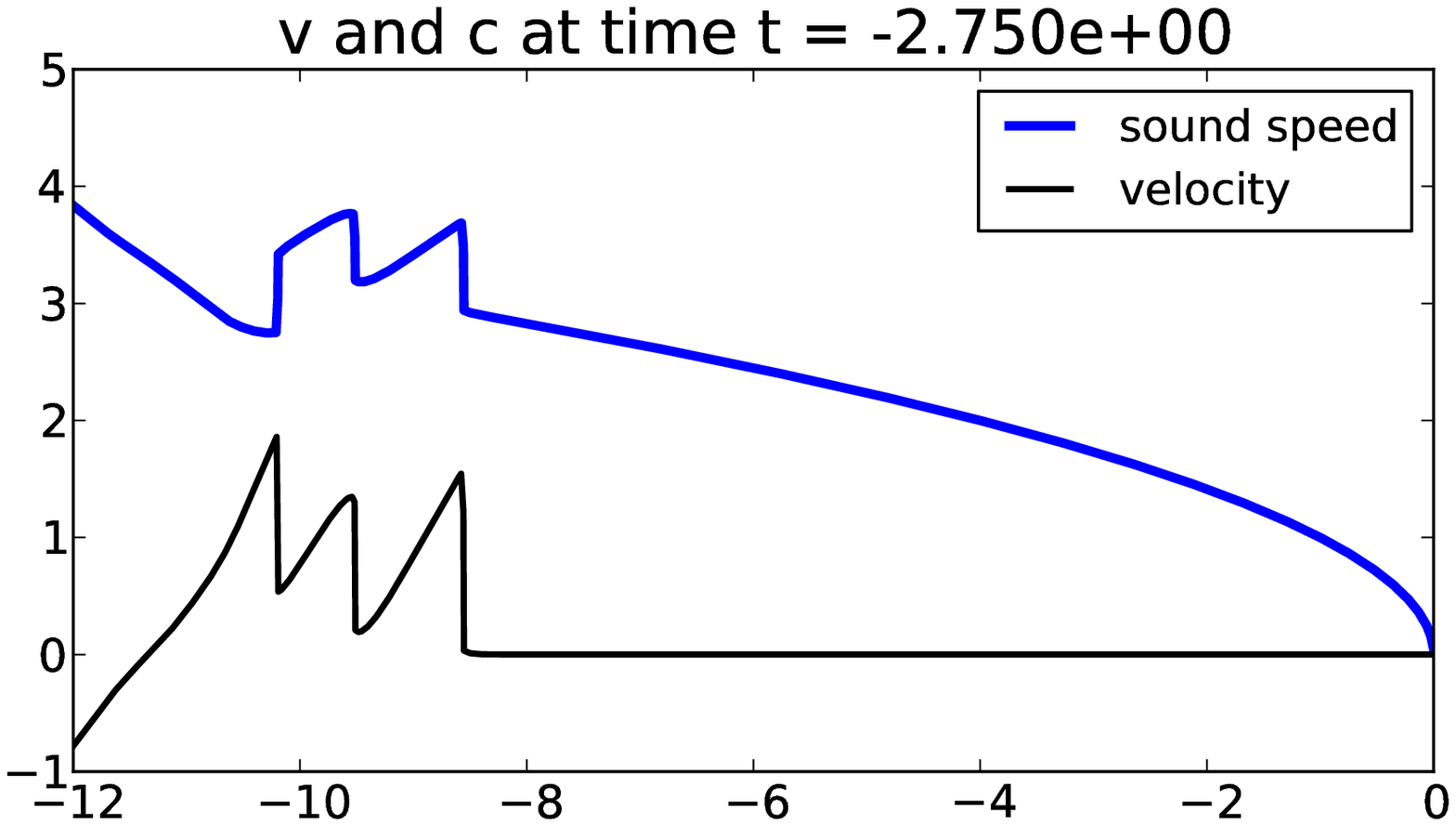} \hfil
\vskip 5pt
\hfil\includegraphics[width=6cm]{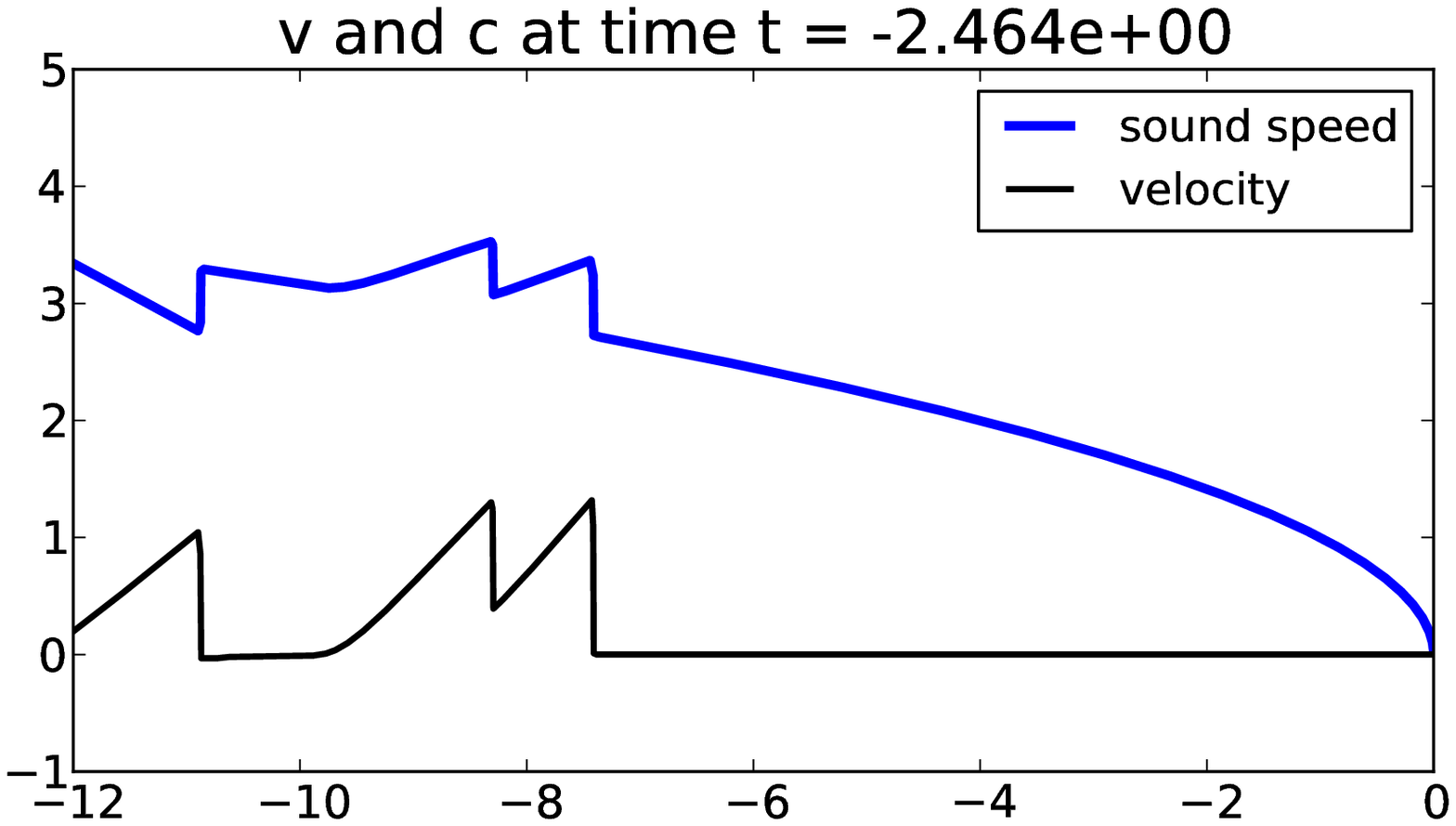}
\hfil\includegraphics[width=6cm]{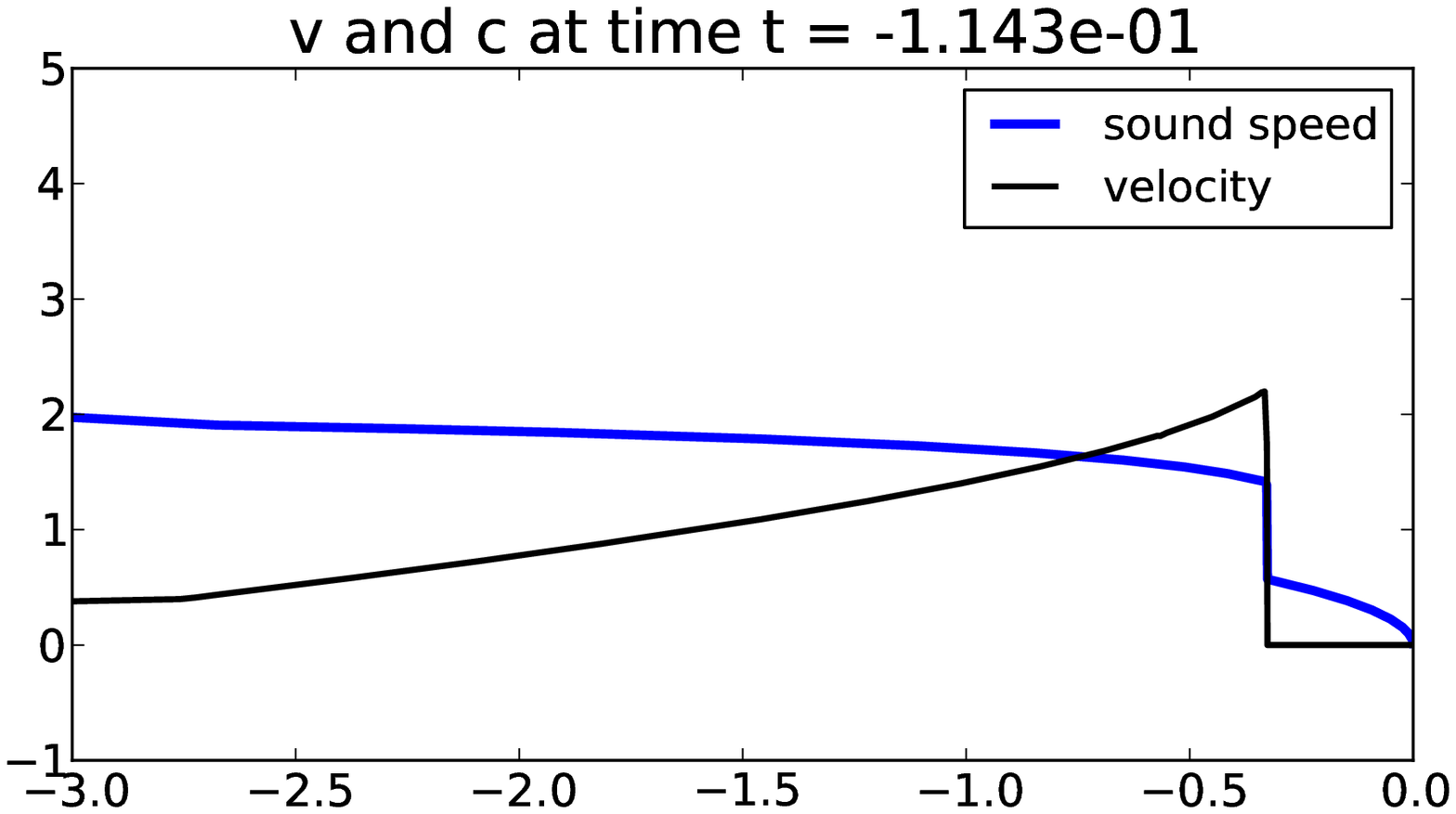} \hfil 
\caption{ 
\label{fig:poly1vc}
Numerical solution for the \new{cold} EOS $n=1.0$, 
showing the fluid velocity and sound speed after 0, 800, 1600, and 8000
time steps on a grid with $M=4800$ grid cells.
The generic Gaussian initial data breaks up into left-going and right-going
waves.  The times listed here and in later figures
are $t_c-t_f$ as discussed in the text,
corresponding to $t$ of the similarity solution.
}
\end{figure}

\begin{figure}
\hfil\includegraphics[width=7cm]{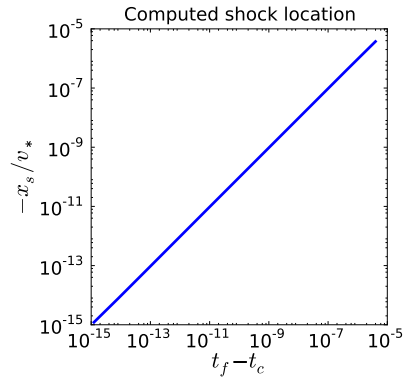} \hfil
\caption{ 
\label{fig:txs}
Computed shock location for the \new{cold EOS} with $n=1$, illustrating
that the shock has constant speed $v_*$ as
$t_c\to t_f$ over many decades.  Results of more than
21000 time steps are plotted (with $M=4800$) on a log-log scale.
}
\end{figure}

\begin{figure}
\includegraphics[width=7cm]{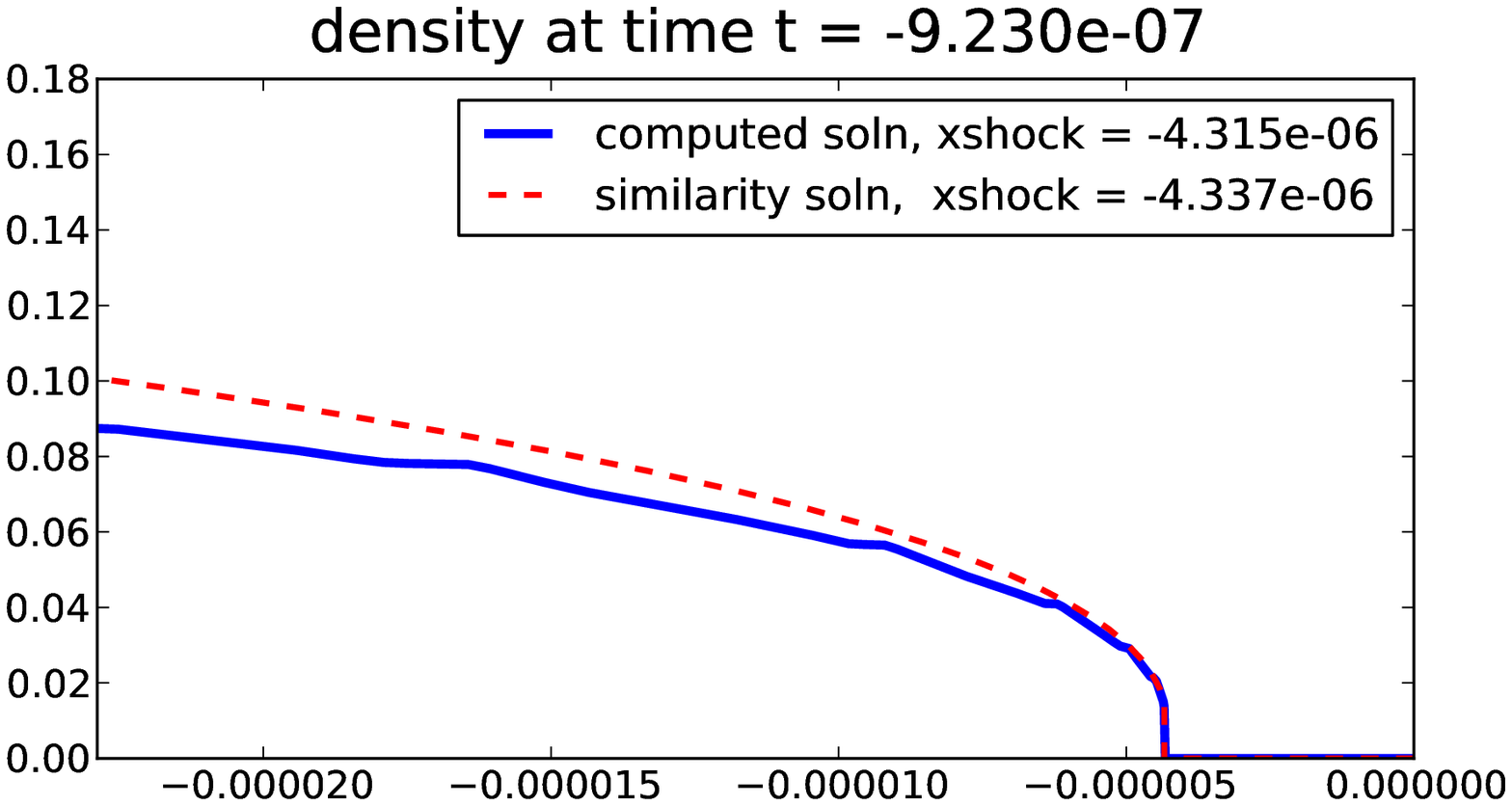} 
\hfil
\includegraphics[width=7cm]{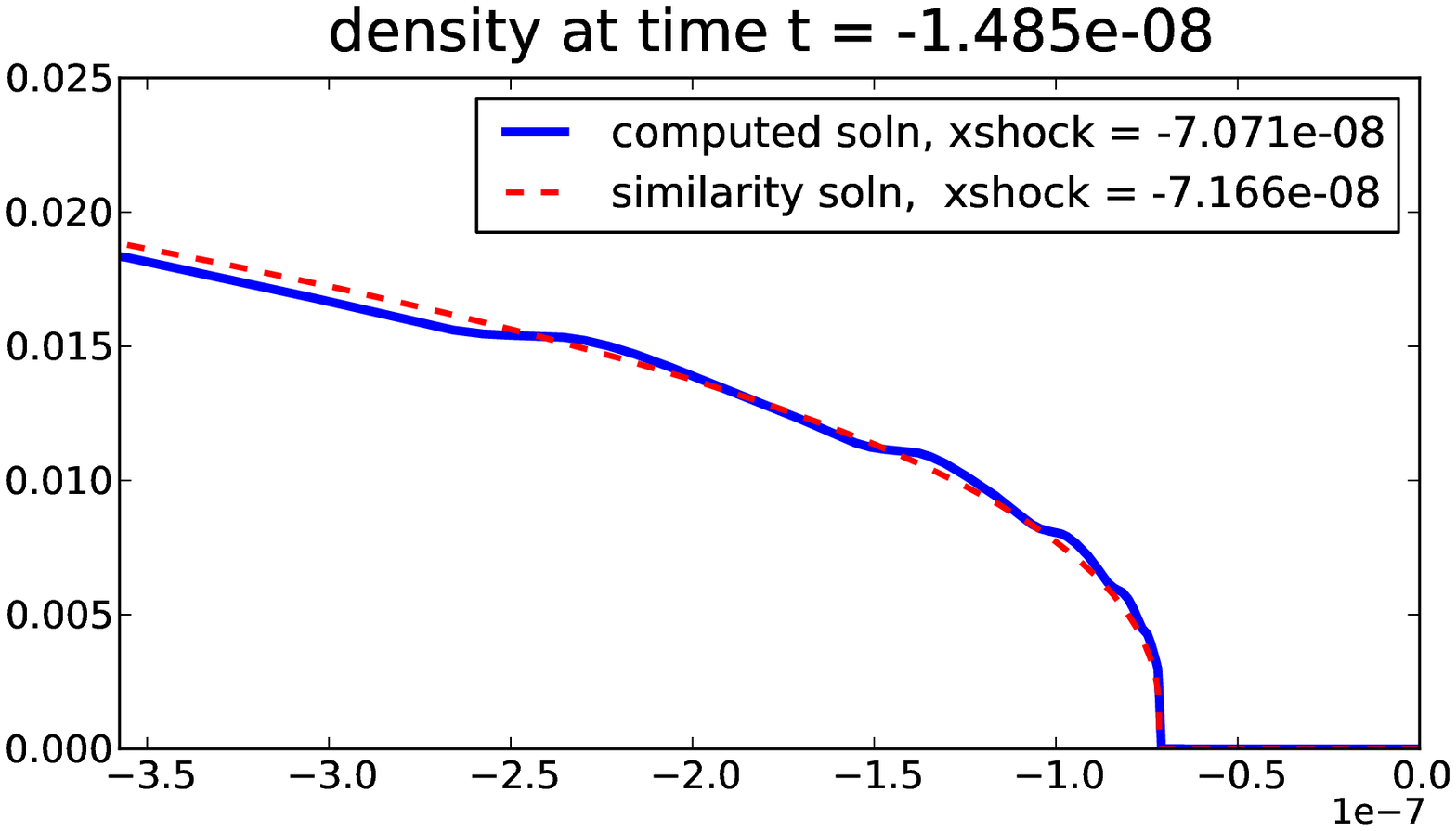} 
\vskip 10pt
\includegraphics[width=7cm]{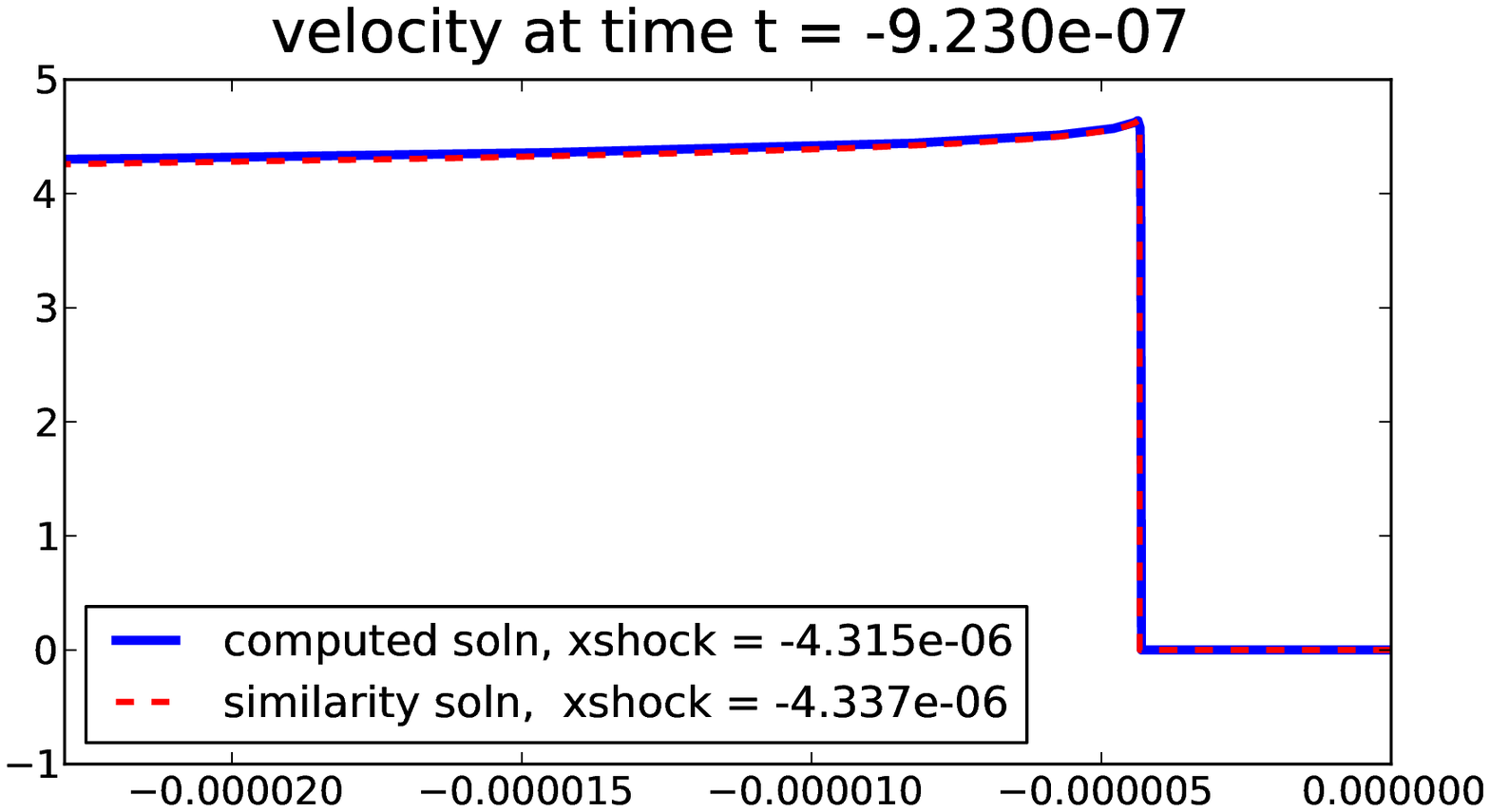} 
\hfil
\includegraphics[width=7cm]{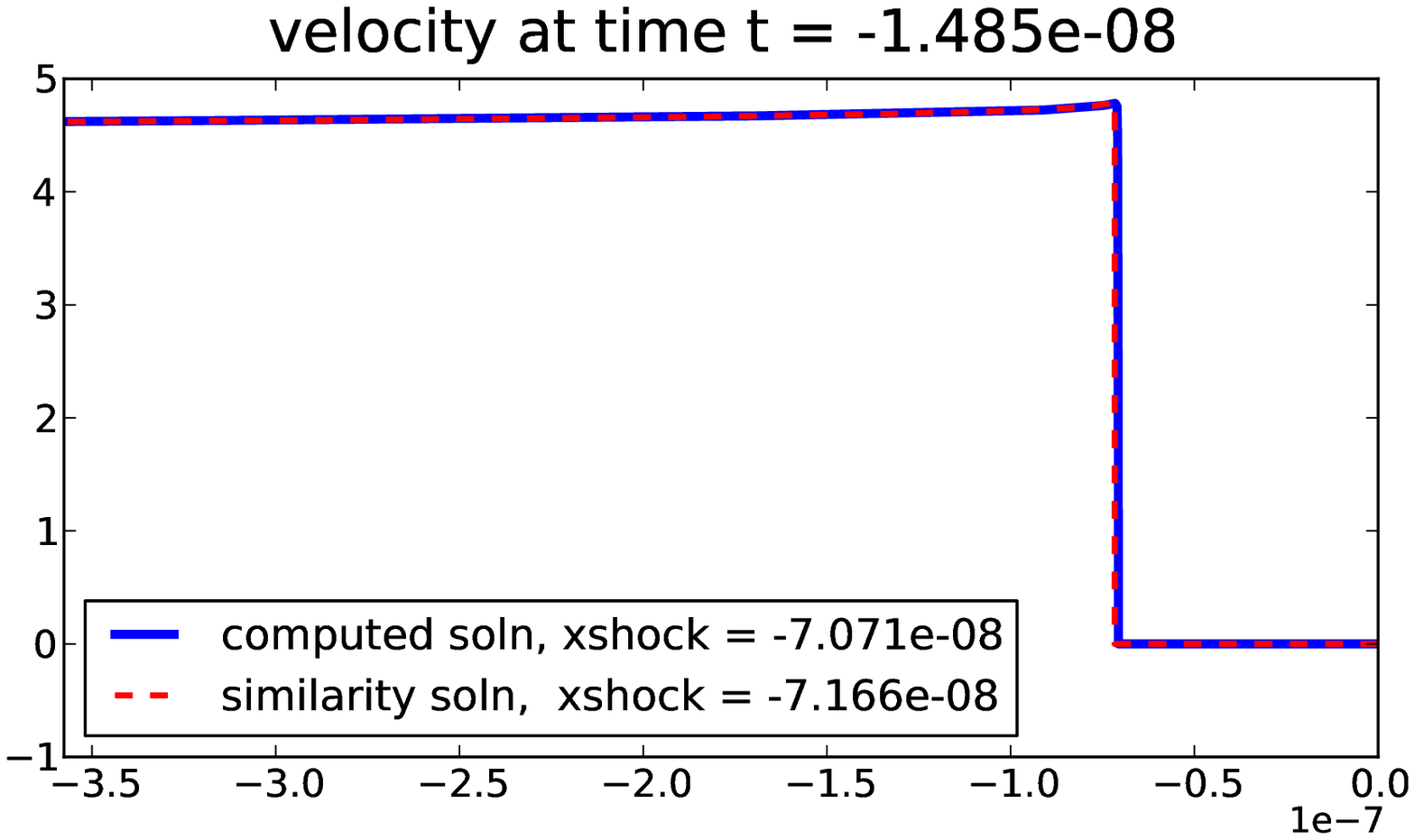} 
\vskip 10pt
\includegraphics[width=7cm]{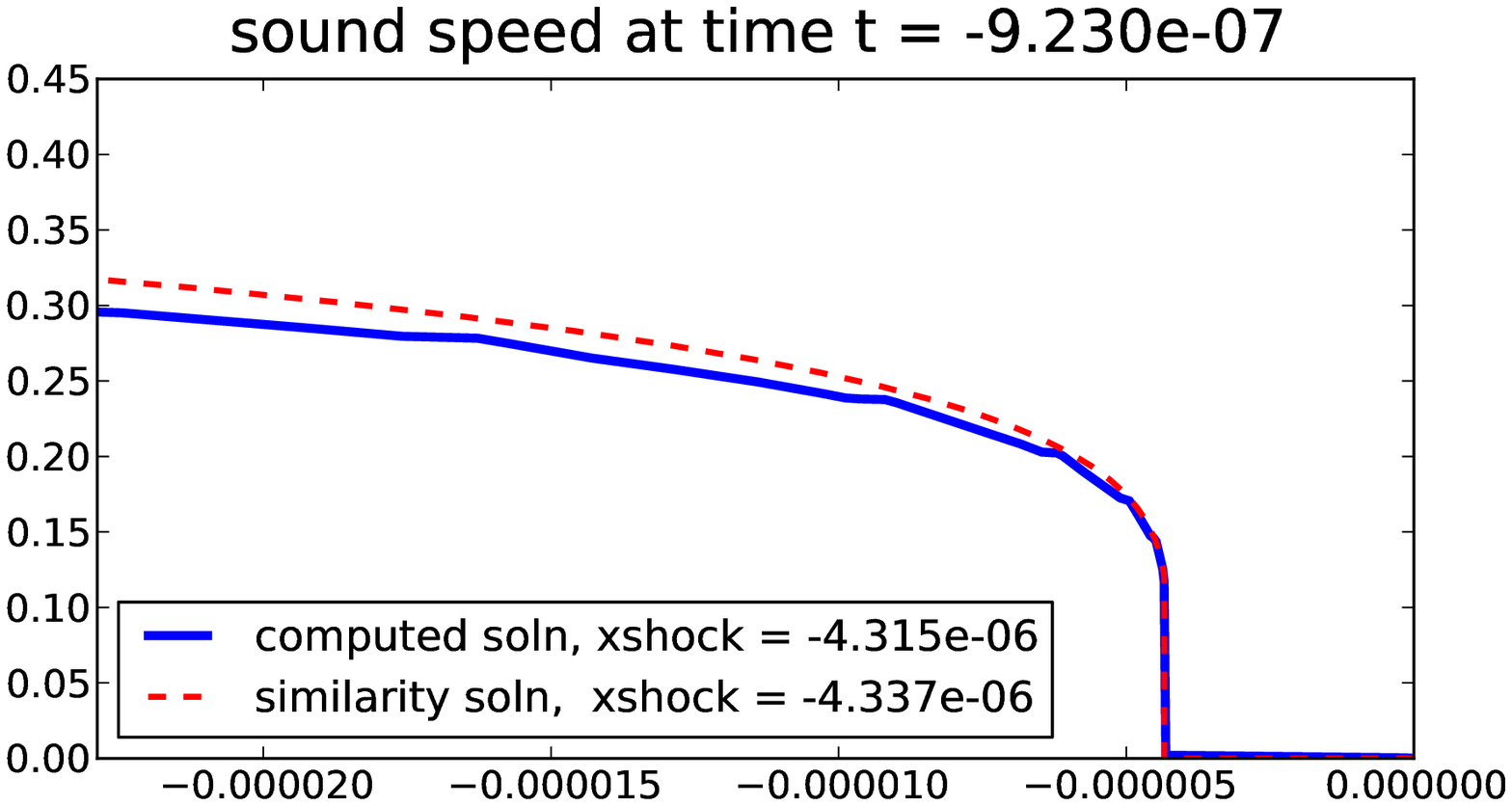} 
\hfil
\includegraphics[width=7cm]{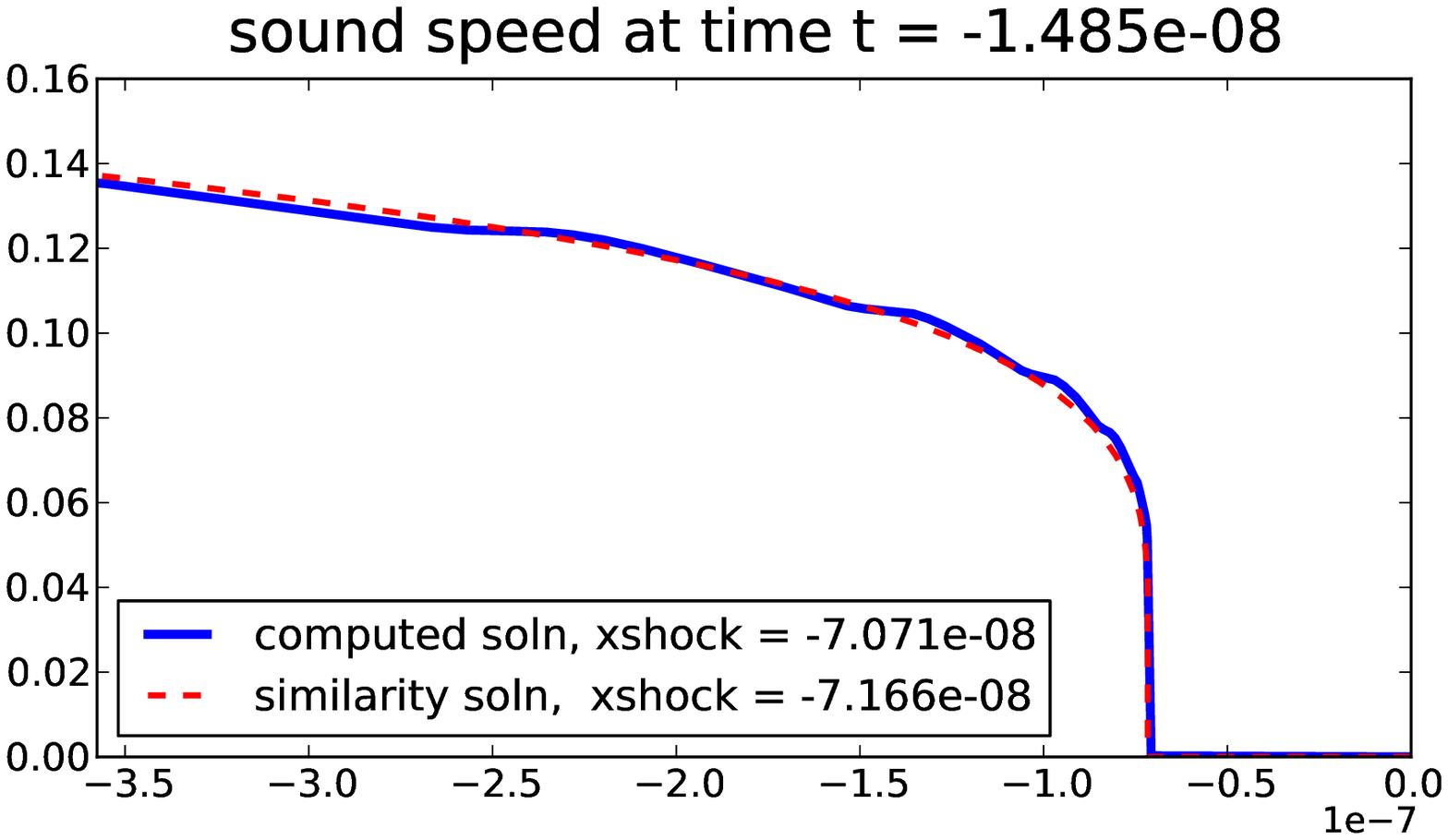} 
\caption{ \label{fig:poly_n1}
\new{Cold} equation of state with $n=1$.
The density (top), velocity (middle), and sound speed (bottom) at two
different times.  The left column shows the solution after 20000 time steps
and the right column shows the solution after 24000 time steps.  In all
cases the solid blue curve is the computed solution with $M=4800$ grid cells
and the red dashed curve is the similarity solution.
For animations, see \cite{cg-rjl:shockvacuum10.web}.
}
\end{figure}

For the \new{cold} EOS with $n=1$, the equilibrium solution has
$\rho(x)=-x$, $v(x)=0$, and $c(x)=\sqrt{-x}$. 
For numerical experiments we
perturbed this with strong generic displacements in density and momentum
centered about different points, namely
\begin{eqnarray}
\rho(x,0) &=&-x+10 \exp\left(-((x+11)/0.3)^2\right), \\
(\rho v)(x,0) &=&30 \exp\left(-((x+10)/0.3)^2\right), 
\end{eqnarray}
on $-12\leq x\leq 0$.  \new{The late-time behaviour is independent of
  these initial data except for the fitting of the free parameters
  $v_*$ and $t_c$, thus showing that the similarity solution is
  universal.}  Figures \ref{fig:poly1vc} through \ref{fig:poly_n1}
show results for this case.

For this case, $M=4800$ grid cells were used and we took 40000 time steps,
at which point the shock was within distance $1\times 10^{-14}$ of the
origin.  This required less than 3 minutes
of CPU time on a MacBook Pro using a single 2.2GHz processor.

A wave travelling right has sharpened into a double N-wave by the time
its leading edge has reached $x=-8.5$. By the time the leading edge
has reached $x=-3.4$, the leading N-wave has been caught up by and
merged with the second N-wave, and the wave going left has steepened
into an N-wave and completely left the grid.

The expected similarity solution is obtained by solving
Eq.~(\ref{exactsolutionc}), multiplied by $y$, numerically for $\bar
c(y)\equiv y\hat c(y)$. 
The parameters $v_*$ and $t_f$ (the final time, when the shock arrives
at the surface) must be fit from the computed solution.
For late times in the computation, the shock appears to propagate at nearly
constant velocity and fitting a linear function to the observed shock
location gives an estimate for $t_f$.  The constant velocity should be
$v_*$, but we have found that a more accurate estimate of $v_*$ is obtained
by examining the 
Riemann invariant $v+2nc=v_*={\rm const}$, which is very nearly constant
behind the shock in the computed solution.

As $x_s\to 0$, the instantaneous profiles of $c$ and $v$ in the
numerical solution behind the shock begin to resemble the similarity
solution first in shape and in the shock location while an overall
factor in $c$ and $v$ agrees only later. By the time the shock is at
$x_s \sim 10^{-3}$, there is agreement within a factor of 2. The
agreement is convincing by the time the shock has reached $x_s \sim
10^{-6}$. This is a strong indication that the similarity solution is
an attractor, at sufficiently small $x_s$ and $t_c-t_f$. The relevant
dimensionless number here is that $x_s$ is about $10^{-6}$ times its
value when a shock first formed from generic initial data. As $x_s\to
0$ as $t_c-t_f\to 0_-$, the agreement is very good at late times, as
illustrated in Figure~\ref{fig:txs}.

Figure \ref{fig:poly_n1} shows the results after 20000 and 24000 time
steps, illustrating close agreement with the similarity solution.  At later
times the agreement is not so good though convergence tests show that at any
fixed time there is increasingly good agreement as the grid is refined.


\subsection{\new{Hot} equation of state}

\begin{figure}
\includegraphics[width=7cm]{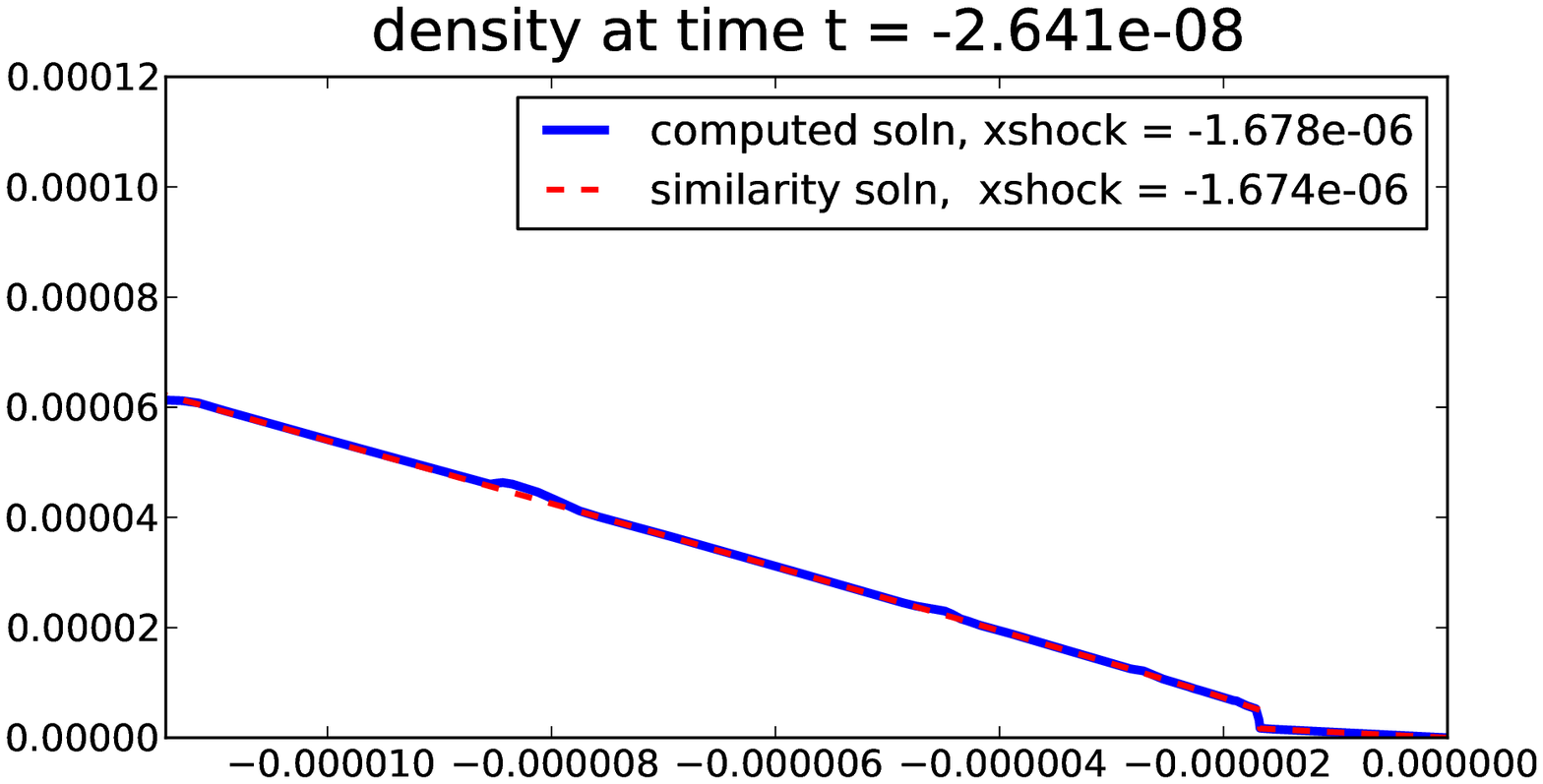} 
\hfil
\includegraphics[width=7cm]{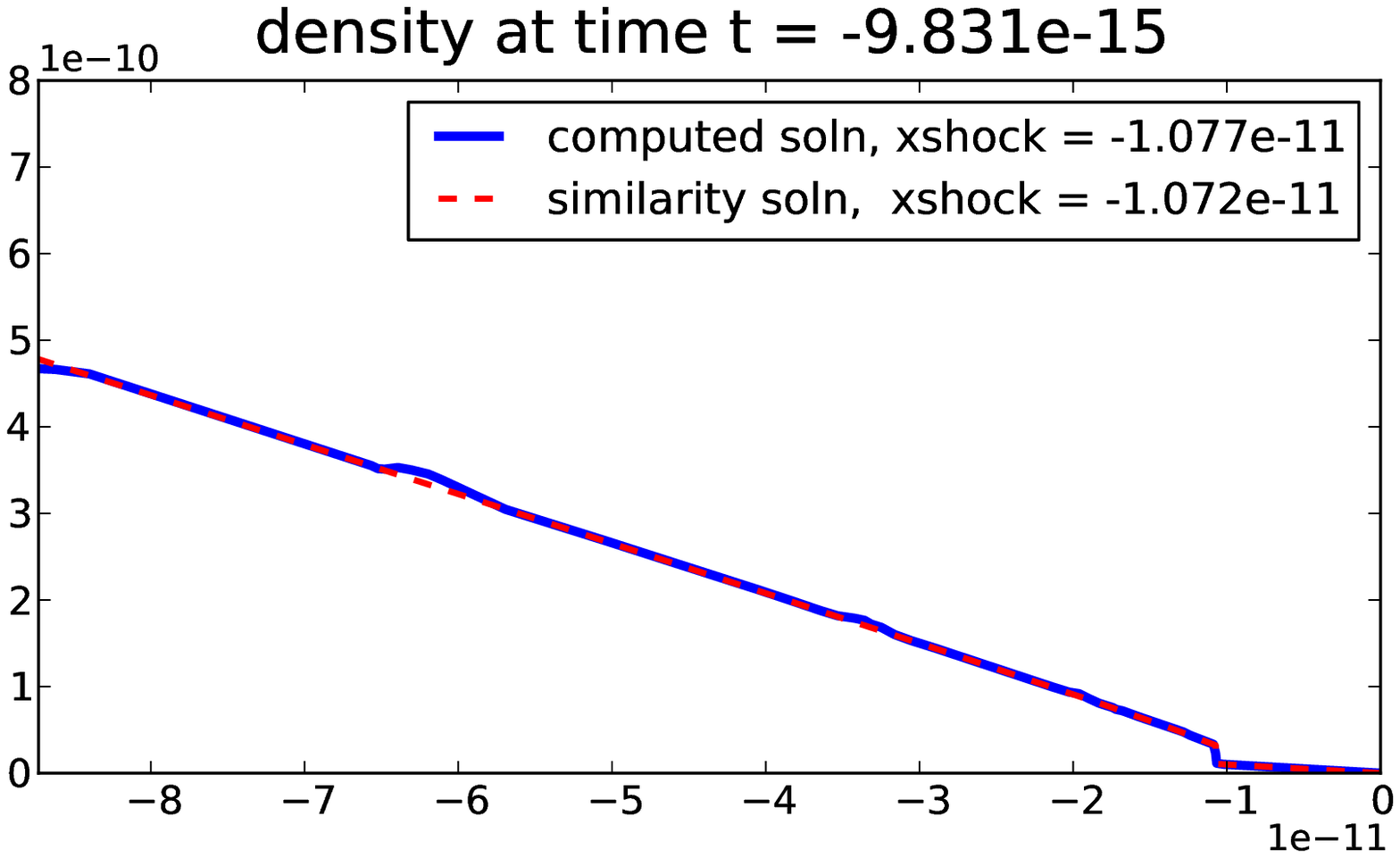} 
\vskip 10pt
\includegraphics[width=7cm]{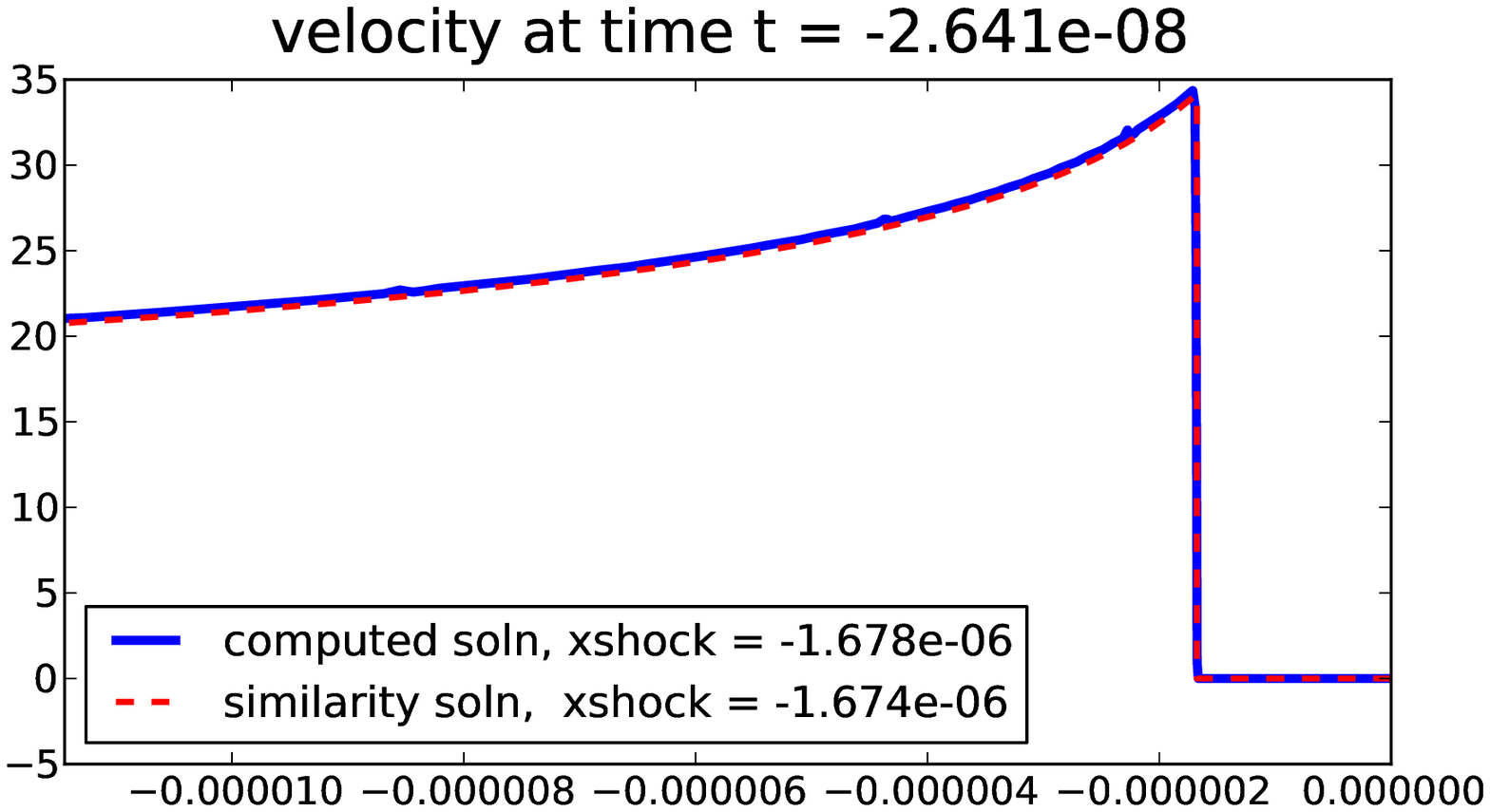} 
\hfil
\includegraphics[width=7cm]{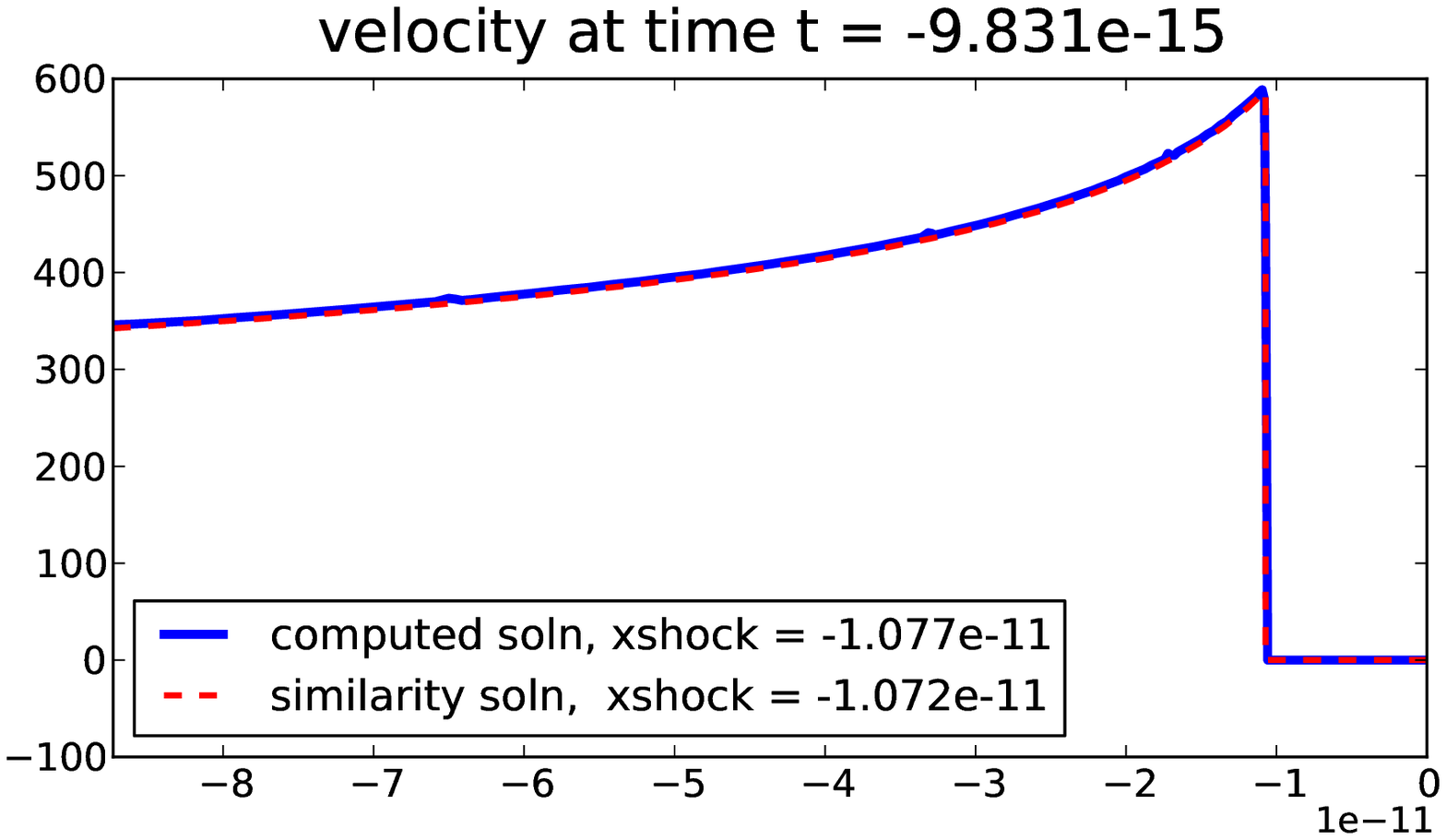} 
\vskip 10pt
\includegraphics[width=7cm]{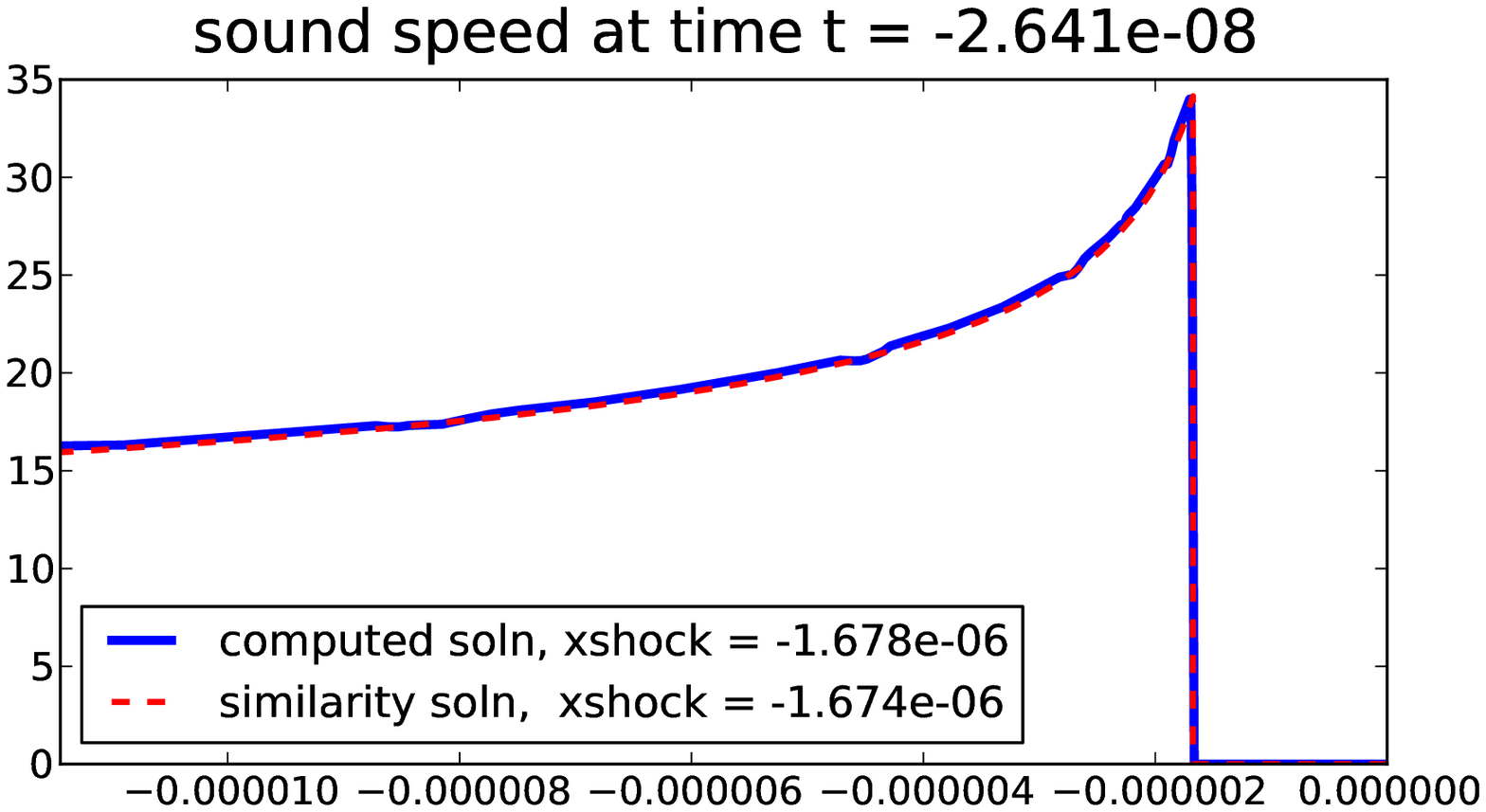} 
\hfil
\includegraphics[width=7cm]{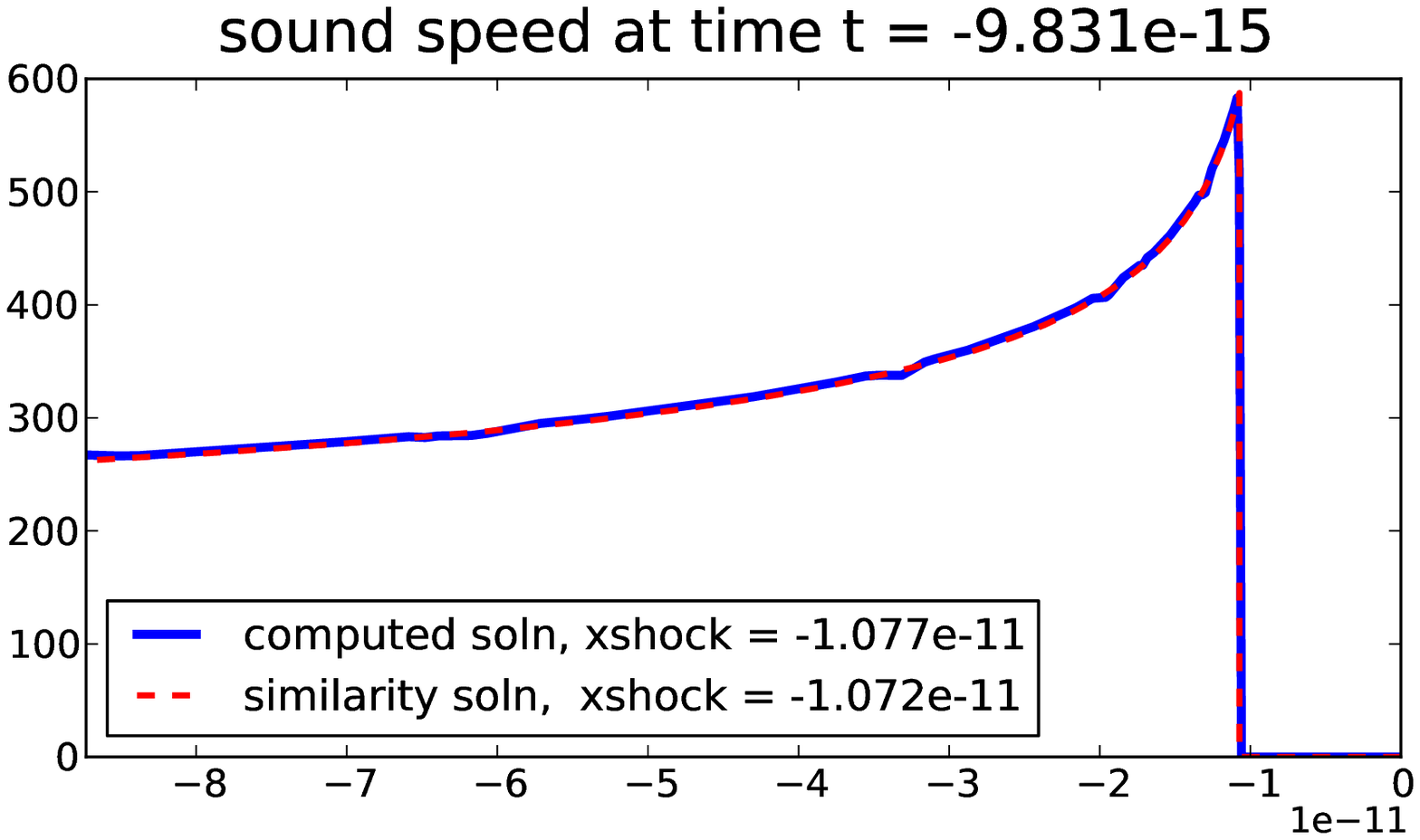} 
\caption{ \label{fig:euler_n1}
\new{Hot} equation of state with $n = n_* = 1$.
The density (top), velocity (middle), and sound speed (bottom) at two
different times.  The left column shows the solution after 6000 time steps
and the right column shows the solution after 10000 time steps.  In all
cases the solid blue curve is the computed solution with $M=1200$ grid cells
and the red dashed curve is the similarity solution.
For animations, see \cite{cg-rjl:shockvacuum10.web}.
}
\end{figure}

\begin{figure}
\includegraphics[width=7cm]{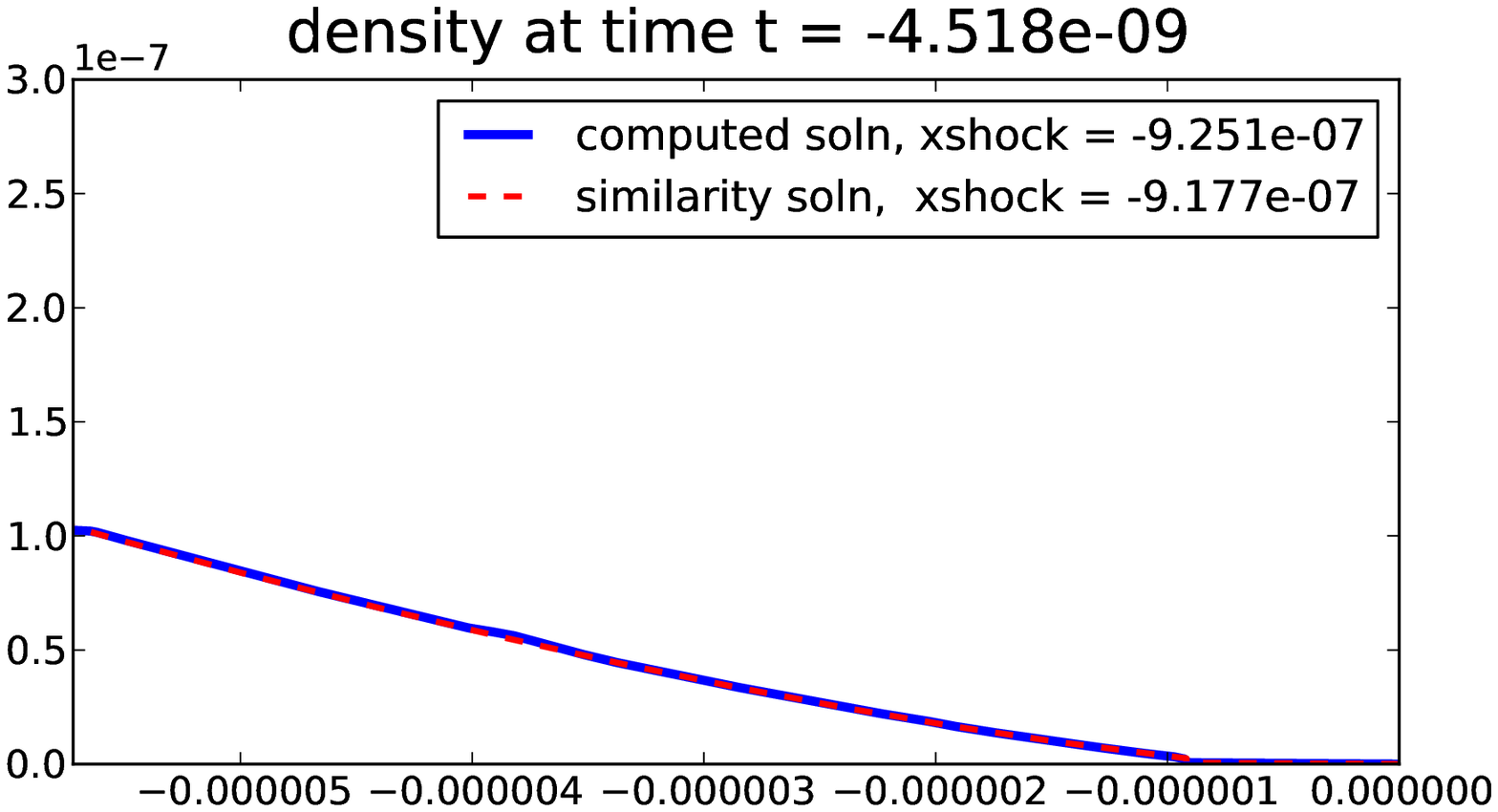} 
\hfil
\includegraphics[width=7cm]{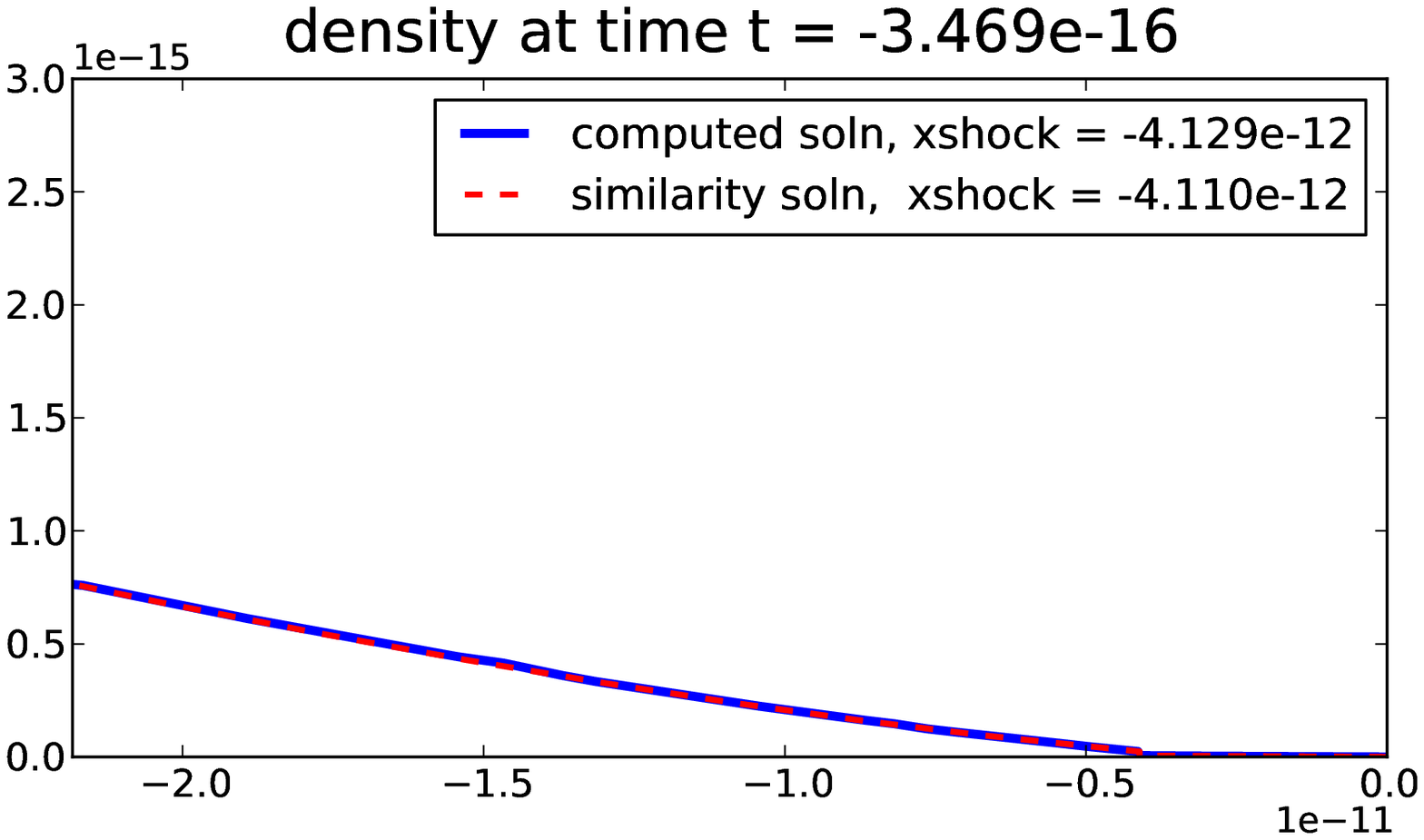} 
\vskip 10pt
\includegraphics[width=7cm]{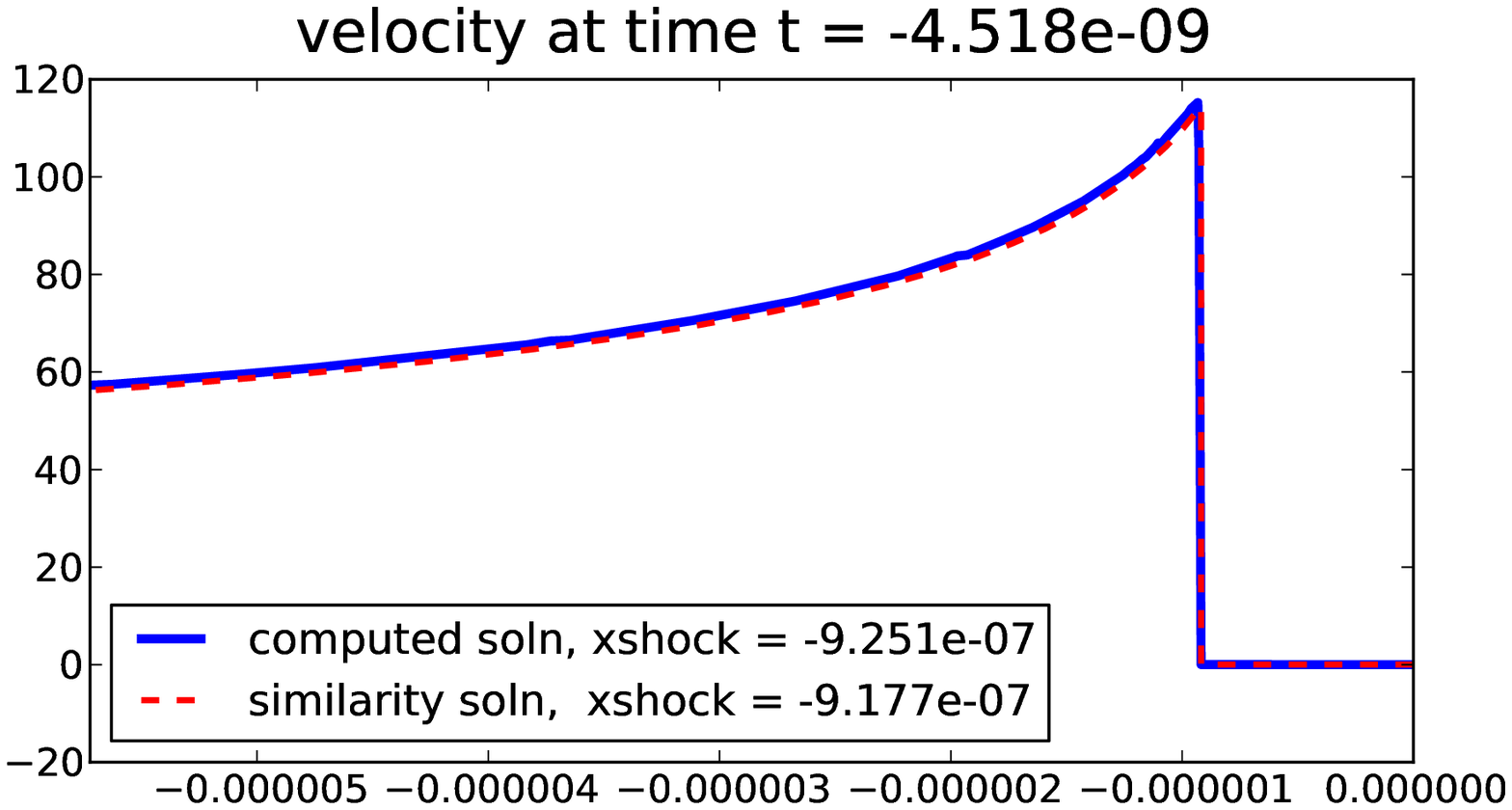} 
\hfil
\includegraphics[width=7cm]{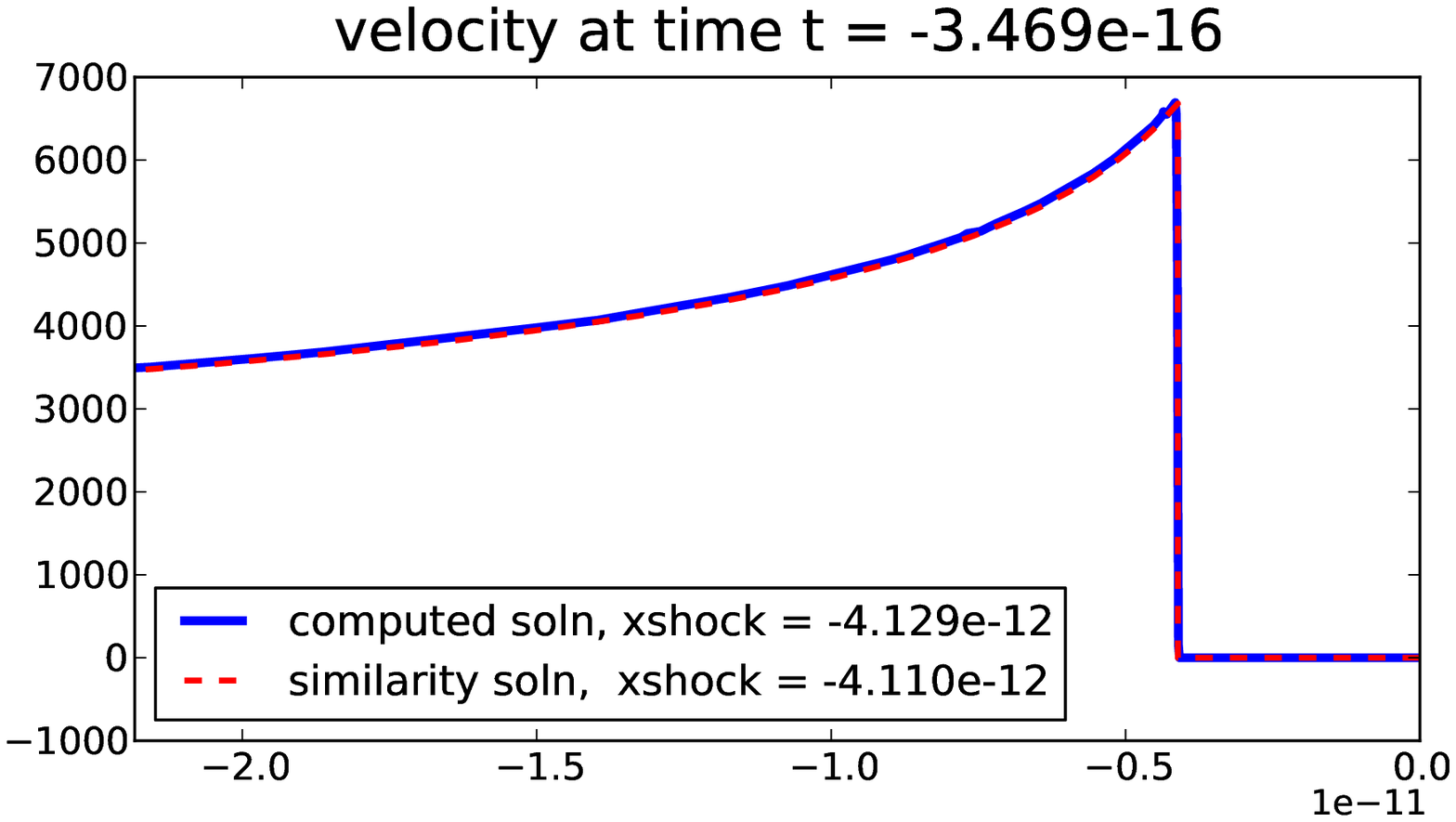} 
\vskip 10pt
\includegraphics[width=7cm]{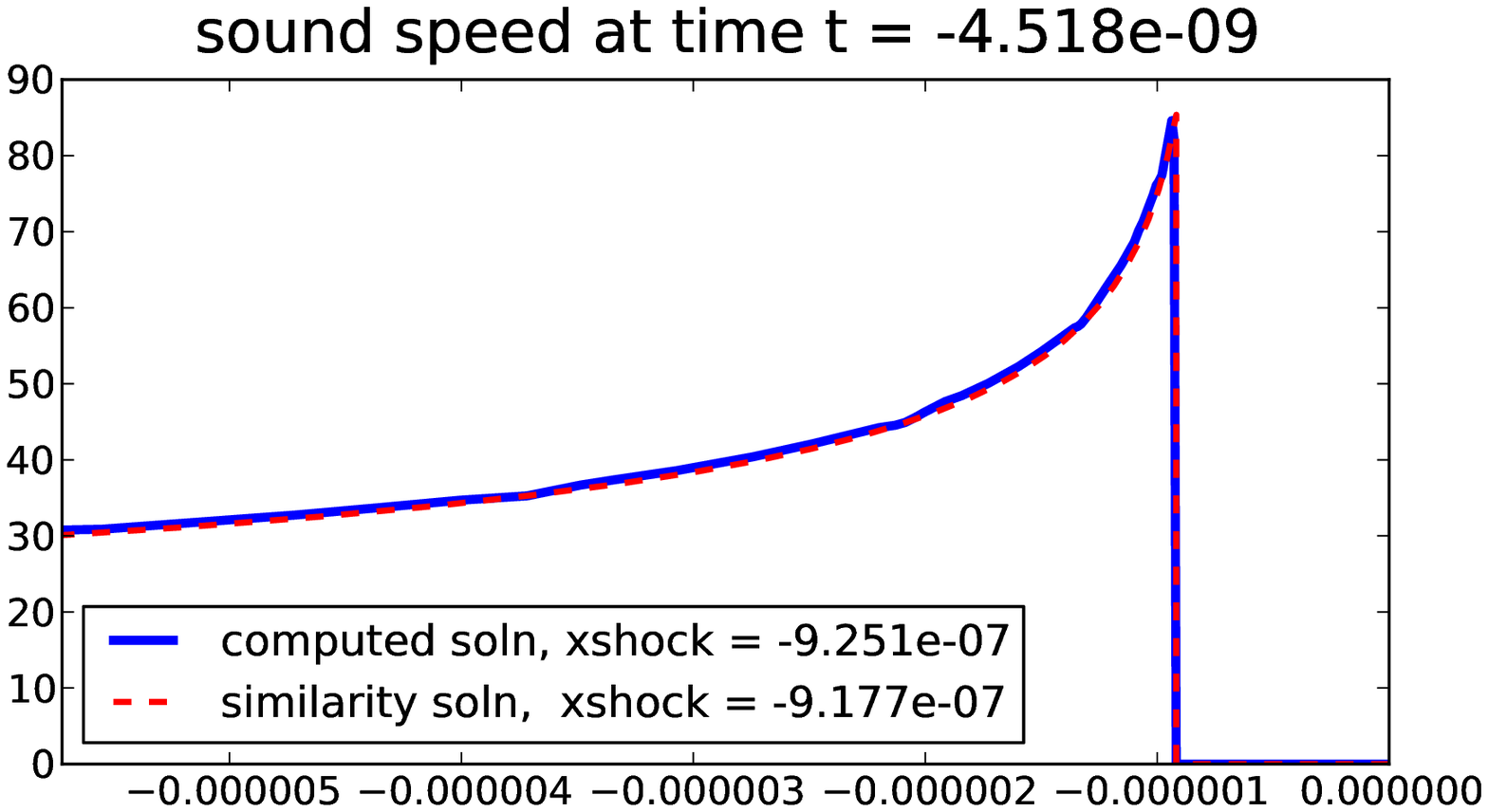} 
\hfil
\includegraphics[width=7cm]{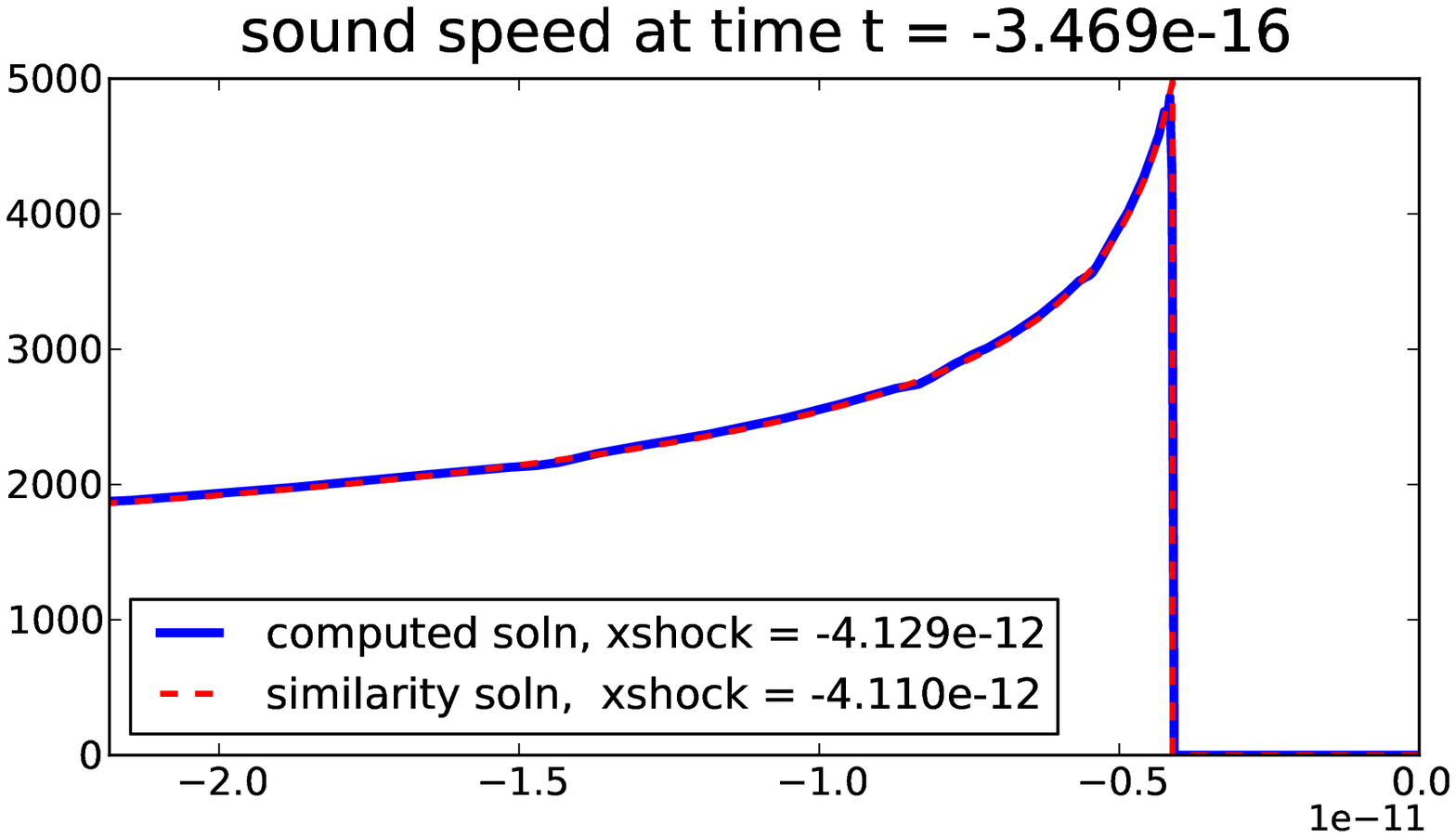} 
\caption{ \label{fig:euler_n1.5}
\new{Hot} equation of state with $n = n_* = 1.5$.
The density (top), velocity (middle), and sound speed (bottom) at two
different times.  The left column shows the solution after 6000 time steps
and the right column shows the solution after 10000 time steps.  In all
cases the solid blue curve is the computed solution with $M=1200$ grid cells
and the red dashed curve is the similarity solution.
For animations, see \cite{cg-rjl:shockvacuum10.web}.
}
\end{figure}

\begin{figure}
\includegraphics[width=7cm]{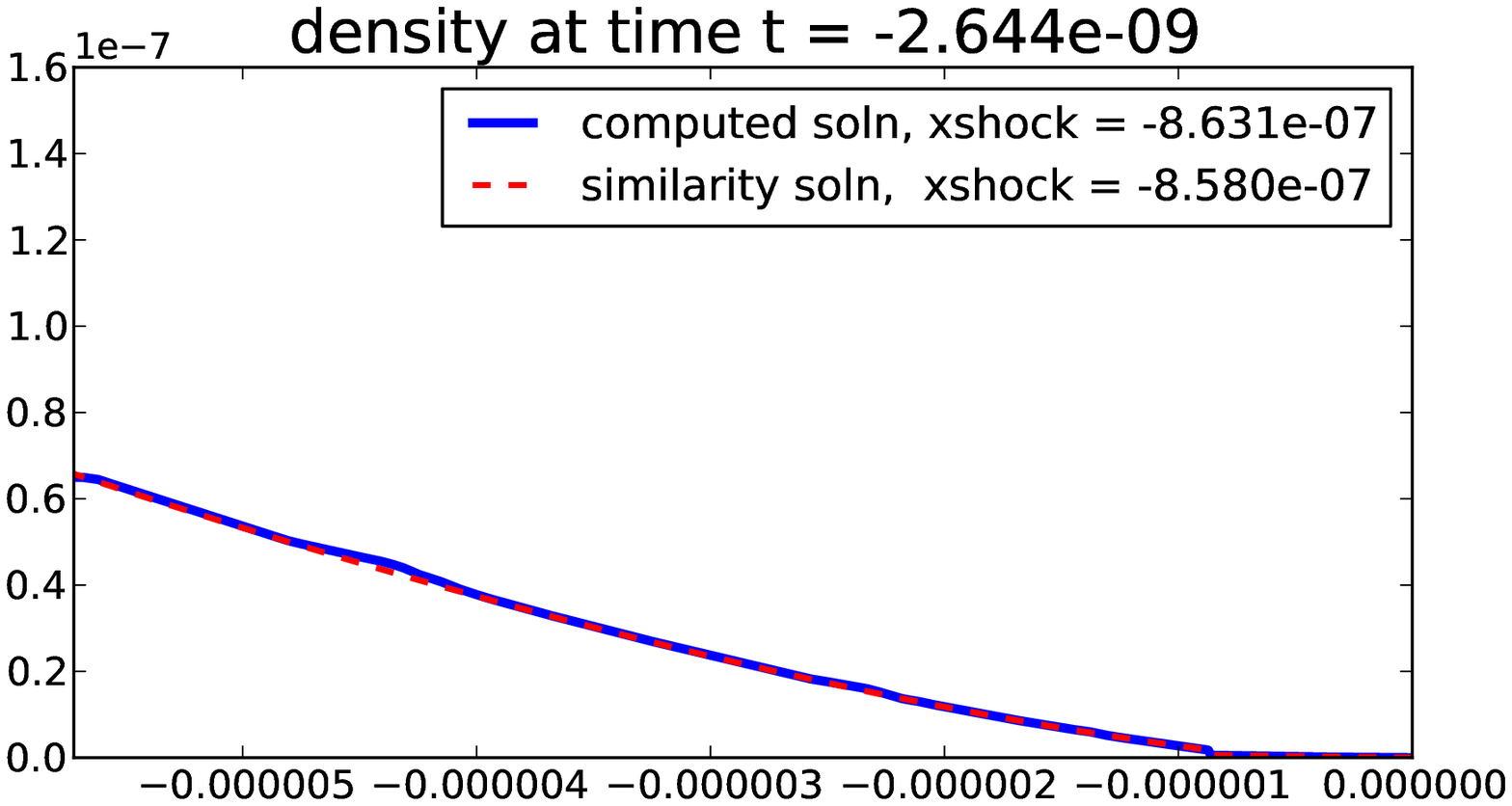} 
\hfil
\includegraphics[width=7cm]{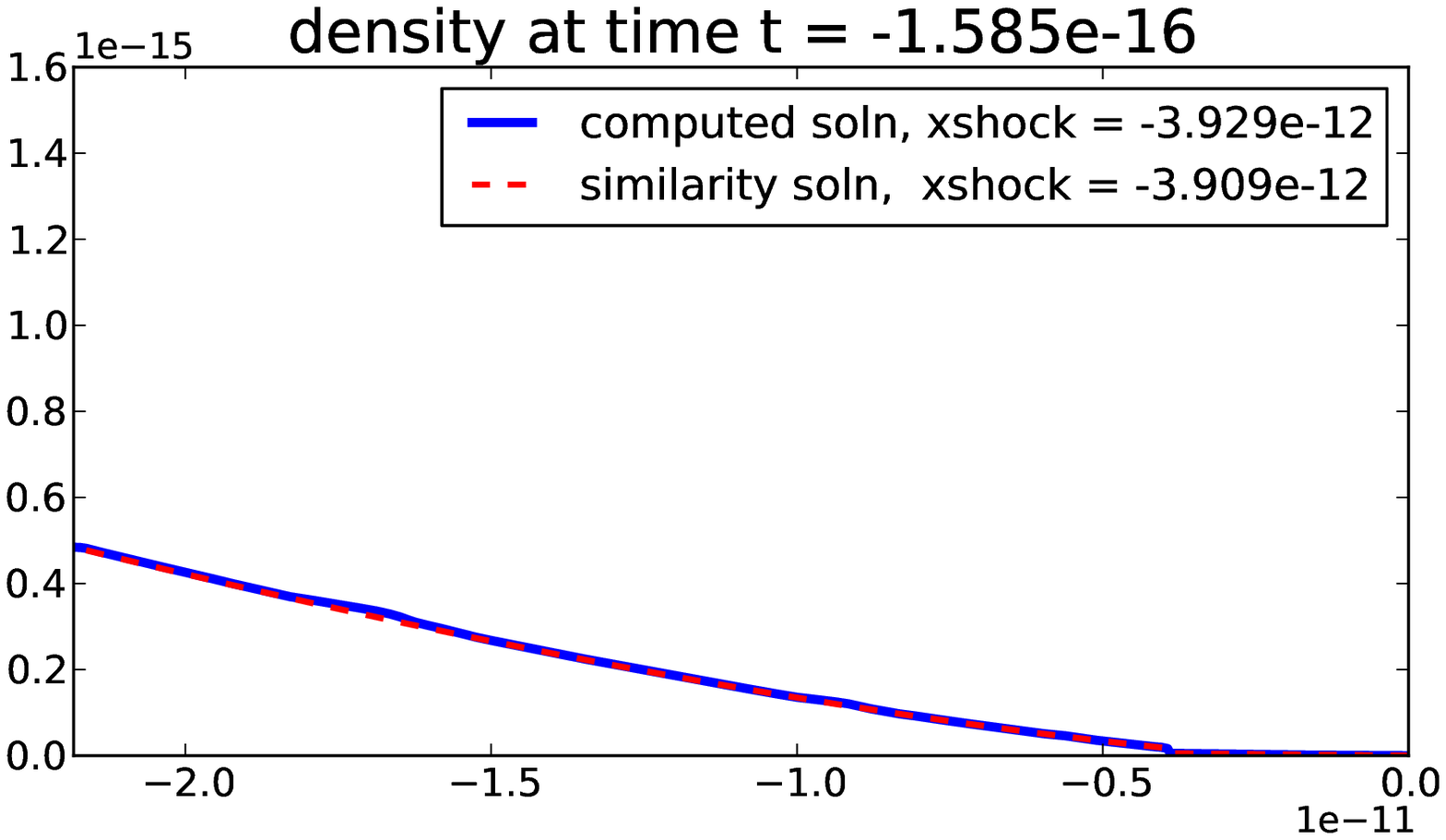} 
\vskip 10pt
\includegraphics[width=7cm]{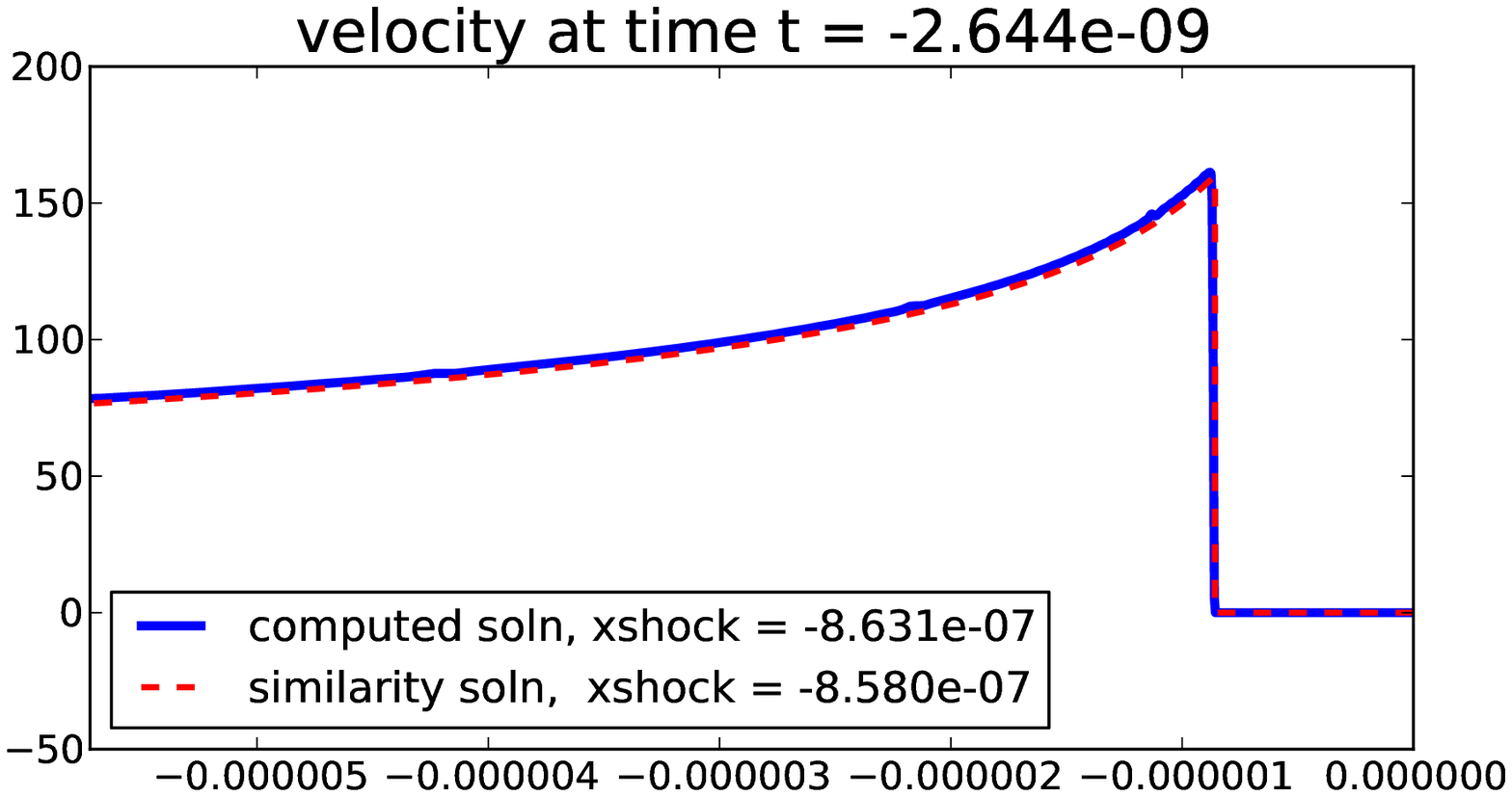} 
\hfil
\includegraphics[width=7cm]{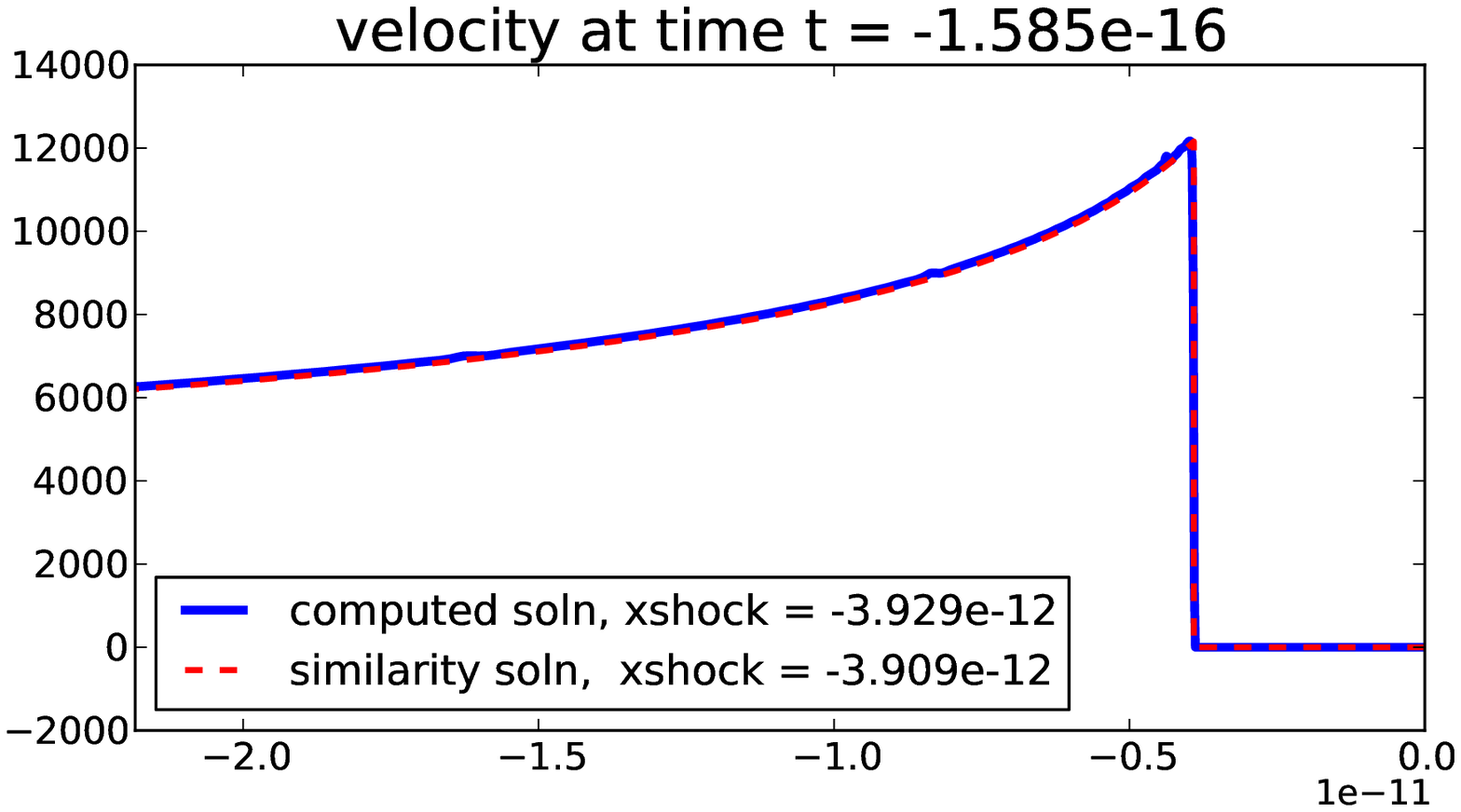} 
\vskip 10pt
\includegraphics[width=7cm]{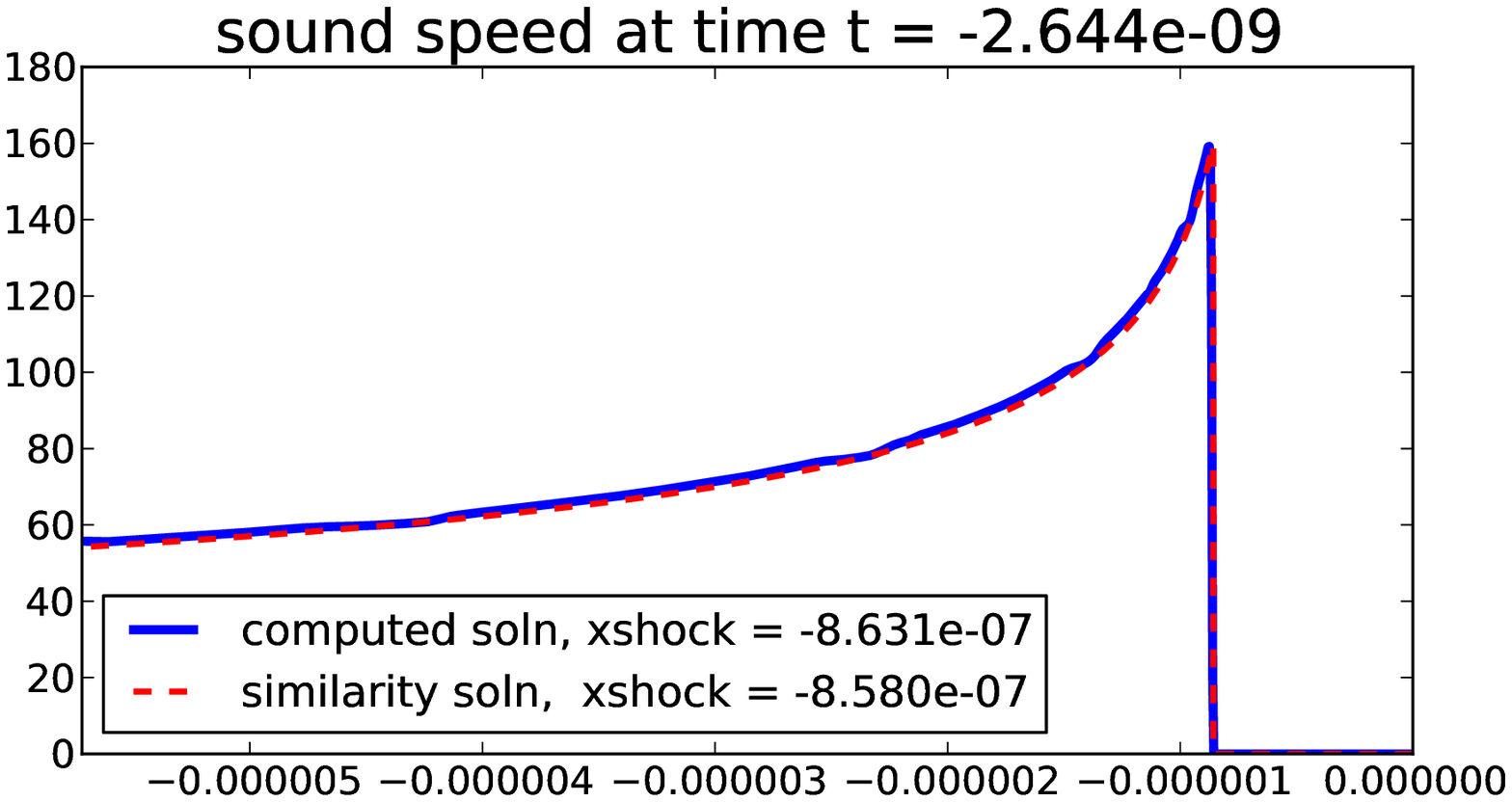} 
\hfil
\includegraphics[width=7cm]{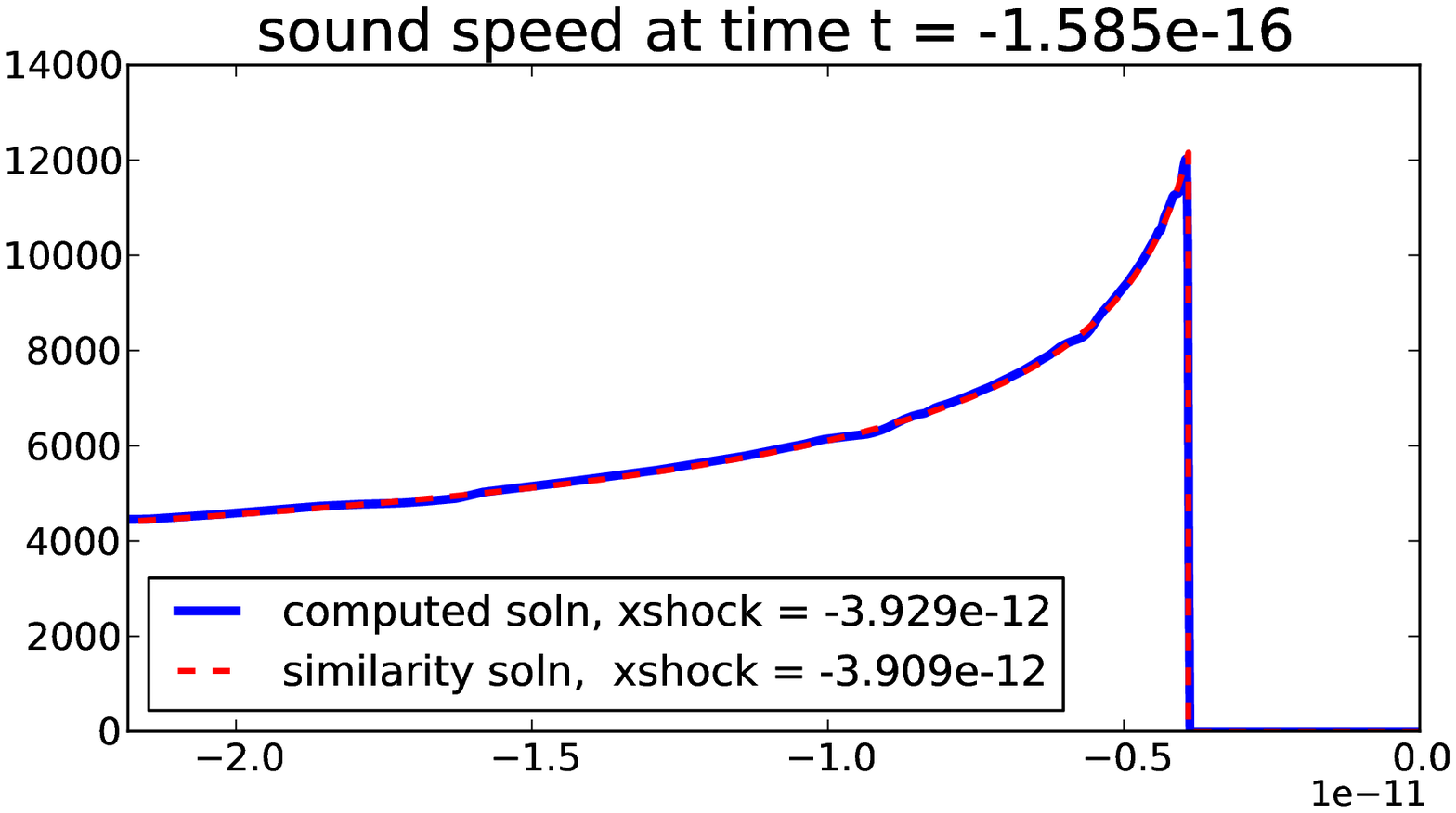} 
\caption{ \label{fig:euler_n1.0nstar1.5}
\new{Hot} equation of state with $n = 1.0,~~n_* = 1.5$.
The density (top), velocity (middle), and sound speed (bottom) at two
different times.  The left column shows the solution after 6000 time steps
and the right column shows the solution after 10000 time steps.  In all
cases the solid blue curve is the computed solution with $M=1200$ grid cells
and the red dashed curve is the similarity solution.
For animations, see \cite{cg-rjl:shockvacuum10.web}.
}
\end{figure}

Figures \ref{fig:euler_n1} through \ref{fig:euler_n1.0nstar1.5} show
results for three \new{hot EOS} cases: $n=n_*=1,~ n=n_*=1.5$, and
$n=1,~n_*=1.5$.  The initial behaviour of the shocks is qualitatively
similar to the \new{cold EOS} case. \new{The late-time behaviour
  approaches the similarity solution and is again independent of the
  initial data except for the fitting of the free parameters $C_y$ and
  $t_c$.}
Note the very different scaling of the axes in each case.

The \new{hot EOS} case is easier to compute than \new{cold EOS} case and all of
these results are shown on grids with $M=1200$ cells.  Finer grids give even
closer agreement. With $M=4800$ the computed solution is
indistinguishable from the
similarity solutions to plotting accuracy.  Moreover, 
the solutions continue to match well out to $x_s\sim 10^{-15}$, after 
starting to agree with the similarity solution very well at $t\sim 10^{-3}$, when 
$x_s\sim 0.01$.

\section{Conclusions}
\label{section:conclusions}

We have constructed similarity solutions that describe a shock
approaching the surface of a star from the interior, in the limit
where the shock is very close to the surface and can be approximated
as planar. We have shown by numerical simulation in the planar
approximation that generic initial data approach the similarity
solution as a universal endstate as the shock approaches the surface. 

The behaviour is very different for cold and hot equations of
state. In the cold EOS, the fluid behind the shock moves to leading
order as a single body, with the matter still in front of the shock
getting stuck on like insects to the front of a car, and dynamically
insignificant. The density directly behind the shock becomes ever
larger than that in front, and the velocity directly behind the shock
approaches a constant.

With the hot equation of state, shock heating becomes essential, and
this indicates that the cold EOS approximation is completely wrong
for shocks near the surface. The ratio of densities in front and
behind the shock approaches a constant. The fluid velocity and sound
speed behind the shock diverge, which means that eventually the motion
has to be treated relativistically. There would then be a
(non-self-similar) transition to a new, ultra-relativistic similarity
solution, where one also has to assume that $P=\rho e/3$. This
ultra-relativistic solution has been investigated by
\cite{NakayamaShigeyama}, but it would depend on the physical
circumstances if it becomes relevant before the perfect fluid
approximation breaks down (and has to be replaced, for example, by
kinetic theory) at the surface of the star.

We hope that our solutions provide at least rough approximations to
shock formation near a stellar surface in some circumstances. They are
also good testbeds for codes capable of dealing with stellar surfaces.
Our simulations are based on the Clawpack software but
required some modification to standard methods in order to compute
into the similarity regime. In particular, we have used a well-balanced
approach that maintains the hydrostatic equilibrium ahead of the shock
to machine precision and a remeshing procedure to obtain well resolved
solutions down to $x_s \sim 10^{-15}$.

In an appendix, we have given a self-contained derivation of the
similarity solution. We stress how similarity solutions for the
cold equation of state can be identified with isentropic
similarity solutions for the hot equation of state, and under what
circumstances any conservation law in the original fluid equations
gives rise to an integral of the motion in the ordinary differential
equations resulting from a similarity ansatz.


\begin{acknowledgments}
CG was supported in part by STFC grant PP/E001025/1. RJL was supported
in part by NSF Grants DMS-0609661 and DMS-0914942, and the Founders
Term Professorship in Applied Mathematics at the University of
Washington.
\end{acknowledgments}


\appendix


\section{General notes on similarity solutions}


\subsection{Transport laws}


Consider a transport law of the form
\begin{equation}
\varphi_t+v\varphi_x=0,
\end{equation}
where $v$ is a velocity, and the dimension of $\varphi$ is
undetermined. Without loss of generality, we can write any power-law
similarity ansatz as
\begin{eqnarray}
\varphi&=&x^\alpha \tilde \varphi(y), \\
v&=&x^\gamma \tilde v(y), 
\end{eqnarray}
where 
\begin{equation}
y\equiv xt^{-\beta}.
\end{equation}
(Any more general ansatz is related to this by redefining $y$, $\tilde
v$ and $\tilde \varphi$.) In order to obtain an ODE in $y$ alone, we
require
\begin{equation}
\label{betagamma}
\gamma=1-{1\over \beta} \Leftrightarrow \beta={1\over 1-\gamma}.
\end{equation}

The resulting ODE can be simplified by defining 
\begin{equation}
\hat v(y)\equiv y^{-{1\over\beta}}\tilde v(y),
\end{equation}
and becomes
\begin{equation}
\left(\hat v-\beta\right)y\tilde\varphi'+\alpha \hat v\tilde \varphi=0.
\end{equation}
This redefinition is equivalent to defining 
\begin{equation}
\label{vhatdef}
v={x\over t}\hat v(y).
\end{equation}
This no longer contains $\gamma$, and makes $\hat v$ dimensionless. 

If, in particular, $\alpha=\gamma$, and we define $\tilde \varphi(y)\equiv
y^{1/\beta} \hat \varphi(y)$, the transport law becomes
\begin{equation}
\label{hathat}
\left(\hat v-\beta\right)y\hat\varphi'+(\hat v-1)\hat \varphi=0.
\end{equation}


\subsection{Conservation laws and integrals}
\label{sec:integrals}


Consider the conservation law
\begin{equation}
\varphi_t+F_x=0.
\end{equation}
Without loss of generality, the most general power law similarity
ansatz that results in an ODE is
\begin{eqnarray}
\varphi&=&x^\alpha \tilde\varphi(y),\\
F&=&x^{\alpha+1}t^{-1}\tilde F(y),
\end{eqnarray}
and this ODE is
\begin{equation}
-\beta \tilde\varphi'+\tilde F'+{\alpha+1\over y}\tilde F=0.
\end{equation}
If and only if $\alpha=-1$, this admits an integral of the form
\begin{equation}
\tilde F-\beta\tilde\varphi=\rm const.
\end{equation}
In particular, if $F=v\varphi$, so that we
have
\begin{equation}
\varphi_t+(v\varphi)_x=0,
\end{equation}
and making the consistent ansatz (\ref{vhatdef}),
we have
\begin{equation}
\left(\hat v-\beta\right)\tilde\varphi=\rm const.
\end{equation}
Usually, however, we cannot arrange for $\alpha=-1$. 

By contrast, if we have a conserved mass density $\rho$, then any quantity
$K$ that is constant along fluid world lines gives rise to an integral
of the ODEs in the self-similarity ansatz. From
\begin{eqnarray}
K_t+vK_x&=&0,\\
\rho_t+(v\rho)_x&=&0,
\end{eqnarray}
we get
\begin{equation}
(\rho K^p)_t+(v\rho K^p)_x=0,
\end{equation}
for any $p$. Assuming
\begin{eqnarray}
\rho&=&x^\alpha \tilde\rho(y), \\
K&=&x^\nu \tilde K(y), 
\end{eqnarray}
and (\ref{vhatdef}), we can now 
choose $p=-(\alpha+1)/\nu$, and obtain
\begin{equation}
\left(\hat v-\beta\right)\tilde\rho \tilde K^p=\rm const.
\end{equation}
without any restriction on $\nu$. 


\section{\new{Cold polytropic gas} dynamics}
\label{appendix:polytropic}


\subsection{Similarity ansatz}


In gas dynamics $u$ and $c$ both scale as velocities. Therefore, with
the ansatz
\begin{eqnarray}
u&=&{\xi\over t}\hat u(y),\\
c&=&{\xi\over t}\hat c(y),
\end{eqnarray}
the equations (\ref{polyuc}) with $g=0$, using (\ref{hathat}), become
\begin{equation}
\label{uhatchatODE}
\left(\hat u\pm \hat c-1\right)\left(\hat u\pm 2n\hat c\right)
+\left(\hat u\pm \hat c-\beta\right)y\left(\hat u\pm 2n\hat
c\right)'=0.
\end{equation}
These equations become automous in the independent variable $\eta=\ln
y$, thus defining a dynamical system in the $\hat u\hat c$-plane. The
direction field is reflection-symmetric about the line $\hat c=0$, and
trajectories cannot cross this line, so that we can assume $\hat c>0$.

We note in passing that a self-similar solution can be obtained also
for $g\ne 0$ if we set $\beta=2$. The dimensionless self-similarity
variable is then $y=x/(gt^2)$, from dimensional analysis
(self-similarity of the first kind).


\subsection{Dynamical system analysis}


The general solution for $\beta=1$ is the trivial one $u=u_0$, $c=c_0$.

For $\beta\ne 1$, the lines $\hat c=\pm (\hat u-\beta)$ define sonic
points, where the lines of constant $y$ are tangential to one of the
two fluid characteristics in the $xt$-plane. The abstract dynamical
system is regular there, with $d\hat c/d\hat u=\mp 1/(2n)$, but the
ODEs in $y$ or $\eta$ are in general singular there, with $d\epsilon/d\eta\sim
1/\epsilon$, where $\epsilon$ is the distance to the line, and so
solution curves end there with $\epsilon\sim\sqrt{\eta_0-\eta}$.

However, on each line of sonic points, there is one regular sonic point given
by $\hat c=\pm \hat u/(2n)$, where $d\epsilon/d\eta$ is finite (because
both its numerator and denominator vanish). They are
\begin{equation}
\hat c=\pm {\beta\over 2n-1}, \qquad \hat u={2n\beta\over 2n-1}.
\end{equation}
Precisely one of these is in the
physical region $\hat c>0$, depending on the sign of $\beta$ and $2n-1$.

There are two smooth solutions through each of these points, with
slopes
\begin{equation}
{d\hat c\over d\hat u}=\mp {1\over2n},\mp{(\beta-1)(4n^2+1)+4n
\over 4n\left[1+2n(\beta-1)\right]}.
\end{equation}

A further power-law expansion shows that for the first of these
solutions $\hat c=\pm 2n\hat u$ to all orders. One of the two ODEs
(\ref{uhatchatODE}) is then identically satisfied, and the remaining
one becomes separable when written in $\hat c$ alone and is
independent of the sign $\pm$ above. An implicit solution in closed
form is given in
(\ref{exactsolutionc}-\ref{exactsolutionu}) above.


\section{\new{Hot polytropic} gas dynamics}
\label{idealgassimilarity}



\subsection{Similarity ansatz}


A useful form of the general power-law similarity ansatz is
\begin{eqnarray}
\rho&=&x^{n_*}\tilde\rho(y),\\
u&=&{x\over t}\hat u(y),\\
c&=&{x\over t}\hat c(y).
\end{eqnarray}
In the variable $\eta=\ln y$ the the resulting ODEs are
autonomous. Furthermore, their right-hand sides are rational functions
of only $\hat u$, $\hat c$ and the parameters ${n_*}$, $\beta$ and
$n$. This means that we have a dynamical system in the $\hat u\hat
c$-plane, with $\ln\rho$ slaved to $\hat u$ and $\hat c$. Again we
assume that $\hat c>0$.


\subsection{Integral of the motion}


The reason that $\tilde\rho$ does not appear on the
right-hand sides is that $K$ is constant on fluid
worldlines, and hence that an integral of the motion exists.
We have
\begin{eqnarray}
\Gamma K&=&c^2\rho^{-{1\over n}}=x^{2-{{n_*}\over n}} t^{-2}\hat c^2
  \tilde\rho^{-{1\over n}}\nonumber \\
\label{Ksimilarity}
&=& x^{2-{{n_*}\over n}-{2\over \beta}} y^{2\over\beta} \hat c^2
  \tilde\rho^{-{1\over n}}
\end{eqnarray}
From our discussion in Sec.~\ref{sec:integrals} we now have the
integral
\begin{equation}
\label{integral}
\left(\hat u-\beta\right)y^{2p\over \beta}\hat
c^{2p}\tilde\rho^{1-{p\over n}}=\rm const.,
\end{equation}
where
\begin{equation}
p\equiv -{{n_*}+1\over q},
\qquad q \equiv 2-{{n_*}\over n}-{2\over \beta}.
\end{equation}


\subsection{Isentropic limit}


To compare the ODEs for $\hat u$ and $\hat c$ with their counterparts
\new{for the cold EOS}, we can write them in the form
\begin{eqnarray}
\label{uhatchatODEidealgas}
&&\left(\hat u\pm \hat c-1\right)\left(\hat u\pm 2n\hat c\right)
+\left(\hat u\pm \hat c-\beta\right)y\left(\hat u\pm 2n\hat c\right)'
\nonumber
\\ &=&
-{n\beta q\over\Gamma}{\hat c^2\over \hat u-\beta}.
\end{eqnarray}
The right-hand side in this equation corresponds to the right-hand
side in Eq.~(\ref{idealalmostRiem}). 

 The \new{cold} similarity equations (\ref{idealalmostRiem}) are
 obtained as the case $q=0$ of the \new{hot EOS} similarity equations
 (\ref{uhatchatODEidealgas}). Conversely, the limit $q\to
 0$ of the integral (\ref{integral}) is
\begin{equation}
\label{isentropicintegral}
y^{2\over \beta}\hat c(y)^2 \tilde\rho(y)^{-{1\over n}}={\rm const}.
\end{equation}
and from (\ref{Ksimilarity}) we see
that $K$ is constant.
Therefore, any combination of ${n_*}$, $\beta$ and $n$ such that
$q=0$ represents an instant of the isentropic limit of the \new{hot EOS}
case, and conversely, any similarity solution \new{with the cold EOS} can be
interpreted as an isentropic similarity solution \new{with the hot EOS}, with
${n_*}$ given by $q=0$, and $\tilde\rho(y)$ given by
(\ref{isentropicintegral}). 


\subsection{Dynamical system analysis}


In general, $\hat u'(\eta)$ and $\hat c'(\eta)$ are singular (their
denominators vanish) on the straight lines $\hat u-\beta=0$ and $\hat
c=\pm (\hat u-\beta)$. These lines divide the half-plane $\hat c>0$
into four sectors. 


\paragraph{Limits $y\to 0$ and $y\to \infty$}


For $|\hat c|, |\hat u|\gg 1,|\beta|$, we have $\hat u,\hat c\sim
y^{-1}$ as $y\to 0$ and and for $|\hat c|, |\hat u|\ll 1,|\beta|$, we
have $\hat u,\hat c\sim y^{-1/\beta}$ as $y\to \infty$. The limit of
$\hat u/\hat c$ is undetermined.


\paragraph{Flow line $\hat u=\beta$}


The line $\hat u=\beta$ represents points where a fluid world line
coincides with a similarity line. Let us call it the flow line. Near
it, $\hat u-\beta \sim \eta-\eta_0$ and $\hat c\sim (\eta
-\eta_0)^{-\beta q/2\Gamma}$.
So solutions end there at finite $\eta$ but in the dynamical system it
is an asymptote.


\paragraph{Sonic lines $\hat c=\pm (\hat u-\beta)$}


The two other lines represent sonic points, where a sound
characteristic coincides with a similarity line of constant $y$. We
shall call them the sonic lines of the dynamical system.  The
dynamical system is regular there, with $d\hat c/d\hat u=\mp
1/(2n)$, but the ODEs in $y$ or $\eta$ are singular there, with
$d\epsilon/d\eta\sim 1/\epsilon$, where $\epsilon$ is the distance to
the line, and so solution curves end there with
$\epsilon\sim\sqrt{\eta_0-\eta}$. On each sonic line, there is also one
regular sonic point where the $y$ or $\eta$ derivatives are also
finite. Exactly one of these is in the physical region $\hat c>0$,
depending on ${n_*}$, $\beta$ and $n$.
Through each regular sonic point there are two smooth solutions.


\subsection{Special case $\beta=1$}


For $\beta=1$, the regular sonic points move down to $\hat c=0$, $\hat
u=1$. There the solution ends because $\hat c$ cannot change sign.

For $\beta=1$ only, the Galileo transformation
$u\to u+v_*$ relates different solutions of the similarity ODEs, so
that any solution $(\hat u(y),\hat v(y))$ can be obtained from a
reference solution $(\hat u_*(y),\hat v_*(y))$ as
\begin{eqnarray}
\hat u(y)-1&=&\left(1-{A\over y}\right)\Bigl[\hat u_*(y-A)-1\Bigr],
\\
\hat c(y)&=&\left(1-{A\over y}\right)\hat c_*(y-A),
\end{eqnarray}
where $A$ is a real constant. Note that this transformation does not
mix solutions in the four sectors, so one needs one reference solution
in each sector to fill the phase plane. 


\bibliographystyle{jfm}
\bibliography{refs}


\end{document}